\documentclass[twocolumn]{aastex631}
\usepackage{amsmath} 
\usepackage{graphicx}
\usepackage{hyperref}
\usepackage[toc,page]{appendix}
\usepackage{cleveref}

\newcommand*{\hi}{\rm{H}\,\rm{\textsc{i}}}
\newcommand*{\msun}{\ensuremath{\rm{M}_{\odot}}}

\newcommand*{\kms}{\text{km}\,\text{s}\ensuremath{^{-1}}}

\begin{document}

\title{SEAMLESS Survey: Four Faint Dwarf Galaxies Tracing Low-Mass Galaxy Evolution Across Environments}

\correspondingauthor{Catherine E. Fielder}
\email{fielder.catherine@gmail.com}

\author[0000-0001-8245-779X]{Catherine E. Fielder}
\affiliation{Steward Observatory, University of Arizona, 933 North Cherry Avenue, Tucson, AZ 85721-0065, USA}

\author[0000-0002-5434-4904]{Michael G. Jones}
\affiliation{Steward Observatory, University of Arizona, 933 North Cherry Avenue, Tucson, AZ 85721-0065, USA}

\affiliation{IPAC, Mail Code 100-22, Caltech, 1200 E. California Blvd., Pasadena, CA 91125, USA}

\author[0000-0003-4102-380X]{David J. Sand}
\affiliation{Steward Observatory, University of Arizona, 933 North Cherry Avenue, Tucson, AZ 85721-0065, USA}

\author[0000-0002-1763-4128]{Denija Crnojevi\'{c}}
\affil{Department of Physics and Astronomy, University of Tampa, 401 West Kennedy Boulevard, Tampa, FL 33606, USA}

\author[0000-0001-9649-4815]{Bur\c{c}in Mutlu-Pakdil}
\affil{Department of Physics and Astronomy, Dartmouth College, Hanover, NH 03755, USA}

\author[0000-0001-8354-7279]{Paul Bennet}
\affiliation{Space Telescope Science Institute, 3700 San Martin Drive, Baltimore, MD 21218, USA}

\author[0000-0001-9775-9029]{Amandine Doliva-Dolinsky}
\affiliation{Department of Physics, University of Surrey, Guildford GU2 7XH, UK}

\author[0000-0001-7618-8212]{Richard Donnerstein}
\affiliation{Steward Observatory, University of Arizona, 933 North Cherry Avenue, Tucson, AZ 85721-0065, USA}

\author[0000-0001-5368-3632]{Laura Congreve Hunter}
\affil{Department of Physics and Astronomy, Dartmouth College, Hanover, NH 03755, USA}

\author[0000-0001-8855-3635]{Ananthan Karunakaran}
\affiliation{Department of Astronomy \& Astrophysics, University of Toronto, Toronto, ON M5S 3H4, Canada}
\affiliation{Dunlap Institute for Astronomy and Astrophysics, University of Toronto, Toronto ON, M5S 3H4, Canada}

\author[0000-0002-7013-4392]{Donghyeon J. Khim}
\affiliation{Steward Observatory, University of Arizona, 933 North Cherry Avenue, Tucson, AZ 85721-0065, USA}

\author[0000-0002-8217-5626]{Deepthi S. Prabhu}
\affiliation{Steward Observatory, University of Arizona, 933 North Cherry Avenue, Tucson, AZ 85721-0065, USA}

\author[0000-0002-0956-7949]{Kristine Spekkens}
\affiliation{Department of Physics, Engineering Physics and Astronomy, Queen’s University, Kingston, ON K7L 3N6, Canada}

\author[0000-0002-5177-727X]{Dennis Zaritsky}
\affiliation{Steward Observatory, University of Arizona, 933 North Cherry Avenue, Tucson, AZ 85721-0065, USA}



\begin{abstract}

We report on four Local Volume dwarf galaxies identified through our ongoing SEmi-Automated Machine LEarning Search for Semi-resolved galaxies (SEAMLESS): Hydrus~A, LEDA~486718, Cetus~B, and Sculptor~26, with the discovery of Hydrus~A reported here for the first time. 
These four galaxies span a wide range of environments and evolutionary states. Hydrus~A ($M_{V}=-9.39\pm0.20$, D $ =3.38^{+0.32}_{-0.30}$~Mpc) and LEDA~486718 ($M_{V}=-11.62\pm0.08$, D $=4.80\pm0.17$ Mpc) are among the most isolated dwarfs known within 5~Mpc, while Cetus~B ($M_{V}=-8.26\pm0.17$, D $=3.32^{+0.25}_{-0.23}$~Mpc) and Sculptor~26 ($M_{V}=-11.25\pm0.10$, D $=3.21\pm0.13$~Mpc) lie $<2 R_{\rm vir}$ of NGC~253. 
Hydrus~A shows properties consistent with quenching driven by cosmic reionization, cosmic-web interactions, or internal feedback. LEDA~486718 is an isolated star forming dwarf. Cetus~B appears quenched and morphologically disturbed, making it a low-mass satellite or backsplash candidate, while Sculptor~26 is red and seemingly gas-poor but displays signs of recent activity, consistent with a transitional evolutionary state. Together, these systems 
demonstrate the power of SEAMLESS for building a census of faint galaxies beyond the Local Group.

\end{abstract}



\section{Introduction} 
\label{sec:intro}

The lowest mass galaxies ($M_{*} \lesssim 10^{6-7}M_{\odot}$) provide stringent tests of galaxy formation physics \citep[e.g.,][]{bullock2017,sales2022}. Much of what we know about dwarf galaxy evolution has come from the Local Group, whose satellites provide unparalleled detail with deep photometric and spectroscopic studies of kinematics and chemistry \citep[e.g.,][]{weisz2014,brown2014,weisz2015,skillman2017,garrison-kimmel2019,nadler2021,akins2021,applebaum2021,manwadkar2022,santos-santos2024,durbin2025,doliva-dolinsky2025}. Yet, this same detail comes at the cost of cosmic context: the Local Group represents a single, environmentally dense pair of massive host galaxies, and may not capture the diversity of dwarf galaxy evolution in the broader Universe.

The Local Group environment 
has significant impacts on dwarf galaxies after infall \citep[e.g.,][]{spekkens2014,putman2021}. The range of environmental processes 
Local Group dwarfs may experience includes tidal interactions and ram-pressure stripping, which can remove gas and truncate star formation \citep[e.g.,][]{mayer2006,slater2013,wetzel2015,akins2021,samuel2022,pathak2025}. Regardless of environment, faint dwarf galaxies also experience the effects of cosmic reionization (by which the least luminous are likely entirely quenched; e.g., \citealt{brown2014,rodriguezwimberly2019}) and internal feedback processes such as those caused by supernovae \citep[e.g.,][]{dekel1986,maclow1999,rey2020}. All of these effects in combination have produced the star formation histories observed among Local Group satellites, making it difficult to disentangle environmental processing from other effects. Studies of dwarfs in other Milky Way–mass environments suggest that the Local Group is not typical: both the abundance and quenching of satellites vary substantially across hosts of similar mass \citep[e.g.,][]{smercina2018,bennet2019,carlsten2022,karunakaran2021,greene2023,karunakaran2023,geha2024,jones2024a,mao2024}. This highlights the need to characterize dwarf galaxies beyond the Local Group — both in isolation, where environmental effects are minimal, and in group environments — in order to capture the full range of evolutionary pathways experienced by low-mass galaxies.

Identification of dwarf galaxies beyond the Local Group does not come easily, as they are inherently faint targets ($M_{V}>-12$) with low surface brightnesses. Until recently, there were only a few known isolated low mass galaxies outside of the Local Group, where dedicated searches were still coming up short \citep[e.g.,][]{sand2015,tollerud2018}. The challenge of weeding out contaminants from genuine dwarf-galaxy candidates using parameter-based cuts has necessitated the adoption of machine-learning approaches to build a broader sample of faint dwarfs outside the Local Group. Our search is specifically designed to identify semi-resolved galaxies: low-mass targets where the brightest stars appear as point sources, in combination with diffuse light from unresolved stars. Semi-resolved targets are likely situated between $\approx1.5-4$~Mpc, based on the tip of the red giant branch standard candle magnitude of $M_{r} = -3.0$ mag \citep{sand2014}.
The SEAMLESS (SEmi-Automated Machine LEarning Search for Semi-resolved galaxies) survey, first presented in \citet{jones2023}, has revealed numerous nearby, low-mass galaxy candidates in Data Release 10 of the DESI Legacy Imaging Surveys \citep{dey2019}. This has already resulted in the discoveries of three isolated dwarf galaxies: Pavo \citep{jones2023}, Corvus~A \citep{jones2024}, and Kamino \citep{mutlupakdil2025}.


In this work, we present deep ground-based follow-up observations of four additional dwarf galaxies identified in SEAMLESS: Hydrus A, Cetus B, Sculptor 26, and LEDA 486718. These systems were selected because they are among the highest-confidence detections in our search. Of this sample, Hydrus~A is newly identified in this work. 
Cetus~B was first reported in the southern SMUDGes (Systematically Mapping Ultra Diffuse Galaxies; \citealt{zaritsky2022}) catalog as SMDG0040349-203322 and presented in the \citet{martinezdelgado2024} search for possible Sculptor group satellites as Do~VIII. Sculptor~26 was first identified in \citet{cote1997} as a potential dwarf galaxy within the Sculptor group, and as a potential low surface brightness dwarf in \citet{karachentsev2000} (as [KKS2000]~02). It is also identified in the LEDA database as LEDA~3097692. LEDA~486718 was identified in the LEDA database and as a dwarf galaxy (AM0316-484) in \citet{arp1987}. Although several of the galaxies studied in this work were previously identified, there was no reliable indication of their distances and thus their physical properties. 
The goal of this work is to establish accurate distances, global physical properties, and environmental contexts for each system. 

In \autoref{sec:data} we discuss the SEAMLESS identification and follow-up deep imaging of these galaxies using the Gemini, Magellan, and Swift facilities. In \autoref{sec:analysis} we present distance and physical property measurements. In \autoref{sec:discussion} we discuss these properties in relation to plausible histories of the galaxies. Last, we summarize and conclude in \autoref{sec:conclusion}.

\section{Observational Data and Reduction} 
\label{sec:data}

Hydrus~A, Cetus~B, Sculptor~26, and LEDA~486718 were identified in our machine learning aided search for semi-resolved galaxies in the publicly available DR10 of the DESI Legacy surveys \citep{dey2019}, which includes the Dark Energy Camera Legacy Survey (DECaLS), the Mayall z-band Legacy Survey (MzLS), and the Beijing–Arizona Sky Survey (BASS). We refer to this program as SEAMLESS, where a discussion of the technique is presented in \citet{jones2023}. Here we briefly summarize.

The SEAMLESS approach to identifying galaxies starts with the SMUDGes pipeline \citep{zaritsky2019,zaritsky2023,zaritsky2025}. High surface brightness stars and bright galaxies are masked and replaced with characteristic noise. Then the images are filtered on a variety of scales in order to enable detection of faint, extended sources. Several criteria \citep{jones2023} and cuts based on \texttt{GALFIT} \citep{peng2002,peng2010} source parameter values are applied in order to mitigate the number of spurious sources. Candidates are then further refined with a convolutional neural network (CNN) classifier. This is the same CNN utilized by SMUDGes, re-trained to identify resolved and semi-resolved low mass galaxies ($D\lesssim4$~Mpc given the DECaLS depth) and to reject cirrus and background galaxies. Sources with a $>90\%$ probability of being resolved or semi-resolved undergo visual inspection by the team in order to identify the most promising candidates for follow-up. 
A few dozen of the highest-confidence candidates are currently undergoing follow-up observations, with four being the focus of this work. These systems are robustly identified in SEAMLESS as high-priority targets, and their diversity in appearance and environment is scientifically interesting, highlighting the breadth of objects our method can uncover. Given the large time demand of follow-up observations, several different facilities have been used in this work.

\subsection{Magellan Megacam}

Deep optical follow-up imaging of Hydrus~A and Cetus~B was obtained with the Megacam imager \citep{megacam} on the Magellan Clay telescope. Megacam has a $\approx$ 24\arcmin $\times $24\arcmin\ field of view and a 0.16 arcsec/pixel binned pixel scale. Observations were acquired on the nights of December 10 (Cetus~B; $0.7''$ seeing) and 11 (Hydrus~A; $0.85''$ seeing), 2023 (part of 2023B-ARIZONA-1; PI: C.~Fielder). Images were taken in the $g$ and $r$ bands, with 14$\times$300~s exposures in $g$ and 12$\times$300~s exposures in $r$ for Hydrus~A in $\approx$ 0.85\arcsec \ seeing and 8$\times$300~s exposures in each filter for Cetus~B in $\approx$ 0.7\arcsec \ seeing. Small dithers were taken between the exposures. 

The Megacam data is reduced with the Megacam pipeline at the Harvard-Smithsonian Center for Astrophysics, which was developed by M. Conroy, J. Roll, and B. McLeod. The reduction includes bias subtraction, flat field corrections, detrending the data, astrometric corrections, and stacking individual dithered frames. After the initial data reduction point-spread function (PSF) fitting photometry is performed on the final stacked images using the {\sc daophot} and {\sc allframe} software \citep{stetson1987,stetson1994} using the procedure of \citet{mutlupakdil2021}. Non-point source objects are removed from the catalogs as outliers in the parameter spaces of $\chi^{2}$ versus magnitude, magnitude error versus magnitude, and sharpness versus magnitude. For point sources that pass these cuts, the instrumental magnitudes are then calibrated to the DECaLS DR10 catalog. These calibrated catalogs then are corrected for Galactic extinction using the \citet{schlegel1998} reddening maps with the \citet{schlafly2011} coefficients. Extinction-corrected magnitudes are used throughout this work.

To determine photometric error and completeness as a function of magnitude and color we perform artificial star tests with the {\sc addstar} routine in {\sc daophot}. Artificial stars are placed into the images on a regular grid spaced $10-20$ times the image FWHM, following the procedure of \citet{mutlupakdil2018,sand2012}. These fake stars are assigned $r$ magnitudes randomly drawn from 18 to 29 mag (favoring fainter stars) and $g$ magnitudes set by a uniform $(g-r)$ color distribution between $-0.5$ and 1.5. Ten iterations are performed per field, yielding $\sim$100,000 stars total. Photometry is derived in the same way as the unaltered image stacks with the same stellar selection criteria applied allowing determination of both the completeness and magnitude uncertainties. The 50\% (90\%) completeness for Hydrus~A is $r = 26.64$ (25.57) and $g = 26.59$ (26.21) mag.  The 50\% (90\%) completeness for Cetus~B is $r = 26.59$ (25.62) and $g = 27.05$ (26.36) mag.

\subsection{Gemini GMOS}

Deep optical follow-up imaging of Sculptor~26 and LEDA~486718 was obtained with the Gemini South telescope using the Gemini  Multi-Object Spectrograph \citep[GMOS;][]{GMOS}. GMOS has an $\approx$ 5.5\arcmin$\times$5.5\arcmin \ field of view and a 0.16 arcsec/pixel binned pixel scale (the same as Megacam). We obtained $g$- and $r$-band imaging under the Fast Turnaround program GS-2024B-FT-208 (PI: M. Jones). LEDA 486718 was observed on November 25 and December 23, 2024, and Sculptor 26 on January 1, 2025. Both galaxies were observed with seven 300-s exposures in each filter. Again, small dithers were taken between the exposures.

We perform initial GMOS data reduction with the {\sc dragons} pipeline \citep{labrie2023,labrie2023b}. Our {\sc dragons} reduction includes bias subtraction, flat-field corrections, bad pixel masking, cosmic ray rejection with the sigma-clipping method, and stacking of the individual dithered frames (weighted average). Then astrometric corrections are applied using {\sc scamp} \citep{scamp}. PSF fitting photometry, Galactic extinction corrections, artificial star tests and photometric error/completeness calculations are then performed in the same manner as that of the Megacam images. The 50\% (90\%) completeness for Sculptor~26 is $r = 26.80$ (25.86) and $g = 27.30$ (26.39) mag. The 50\% (90\%) completeness for LEDA~486718 is $r = 27.05$ (26.21) and $g = 27.60$ (26.56) mag.

\begin{figure*}
    \centering
    \includegraphics[width=1.0\columnwidth]{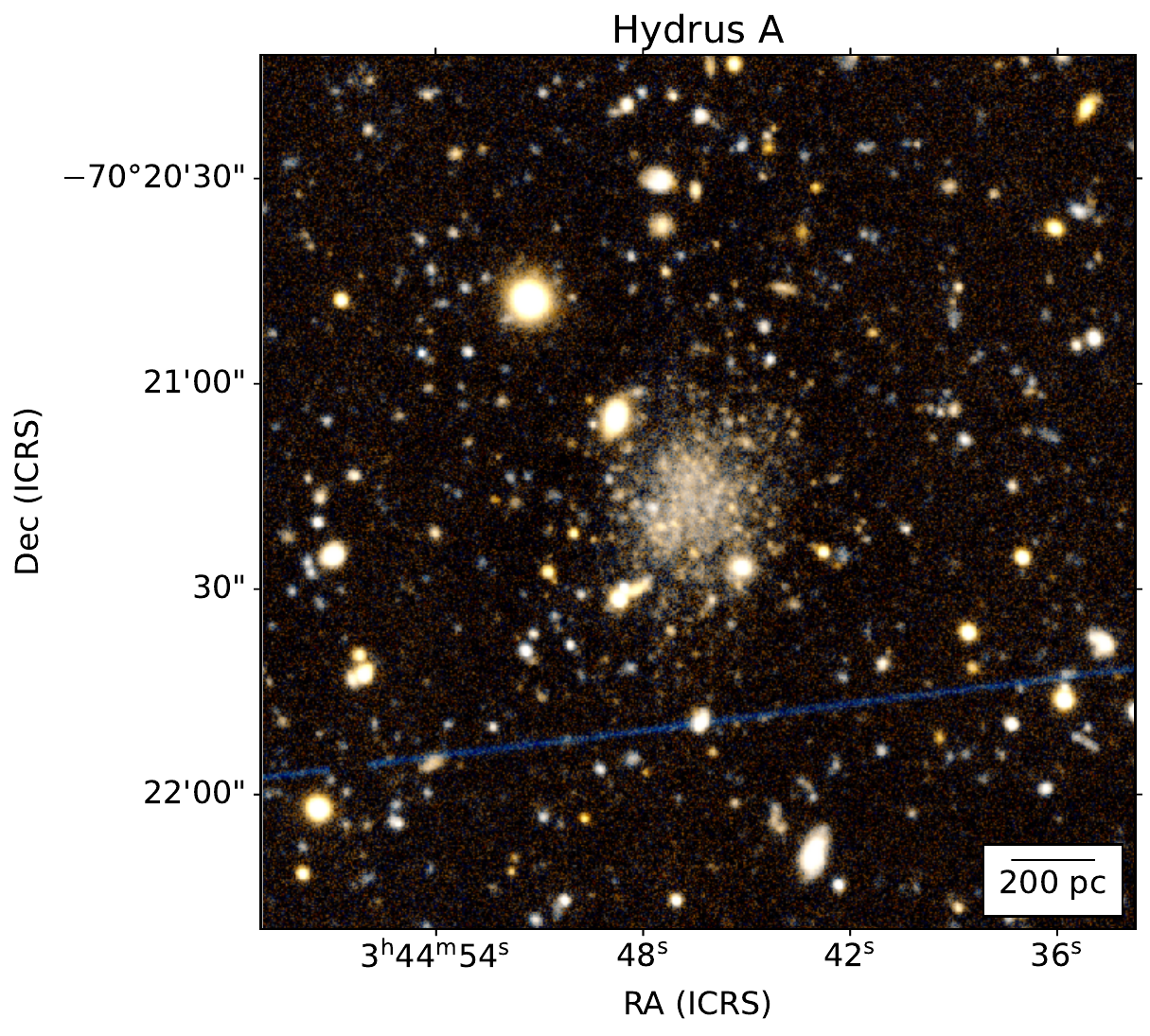}
    \includegraphics[width=1.0\columnwidth]{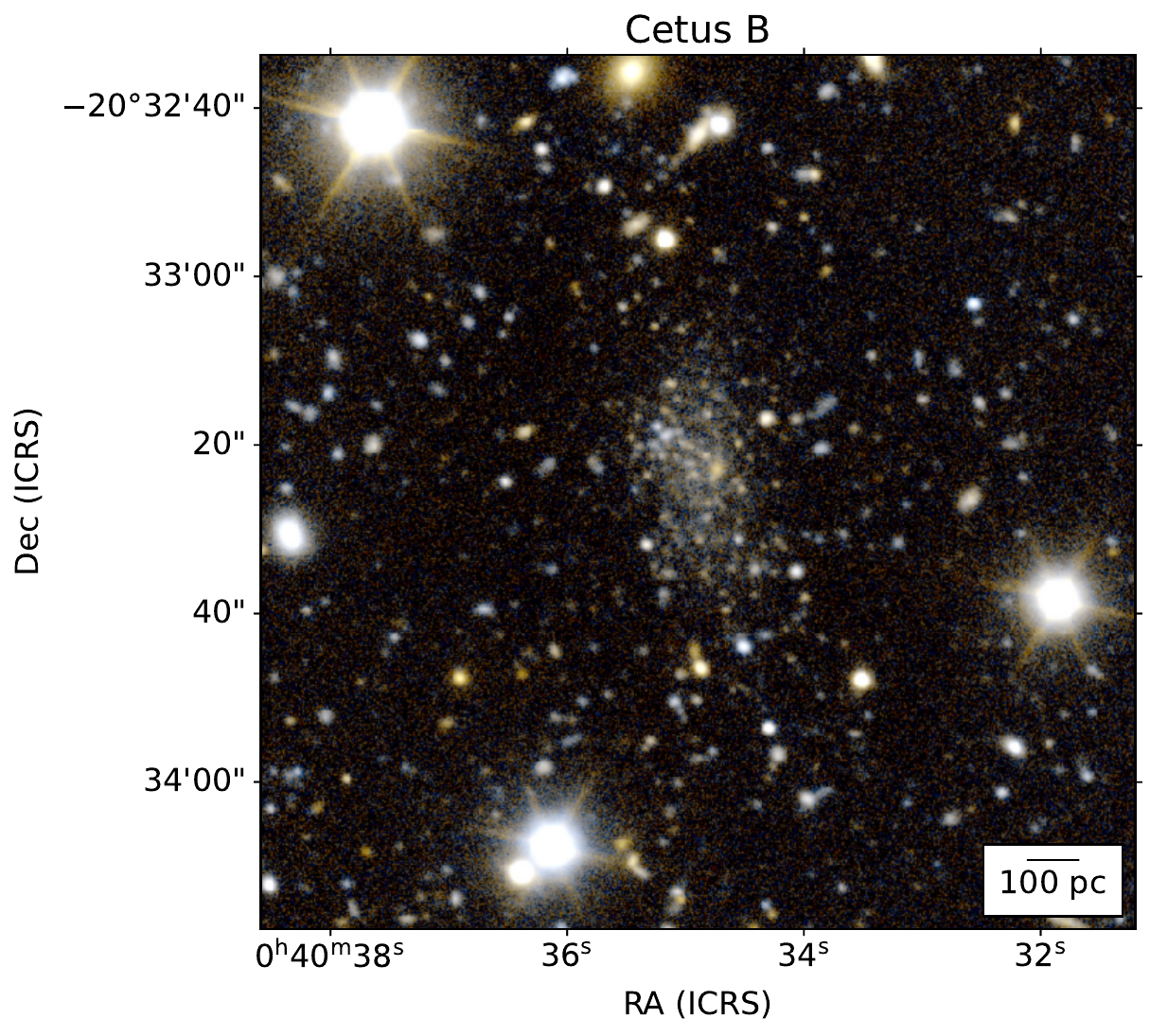}\\
    \includegraphics[width=1.0\columnwidth]{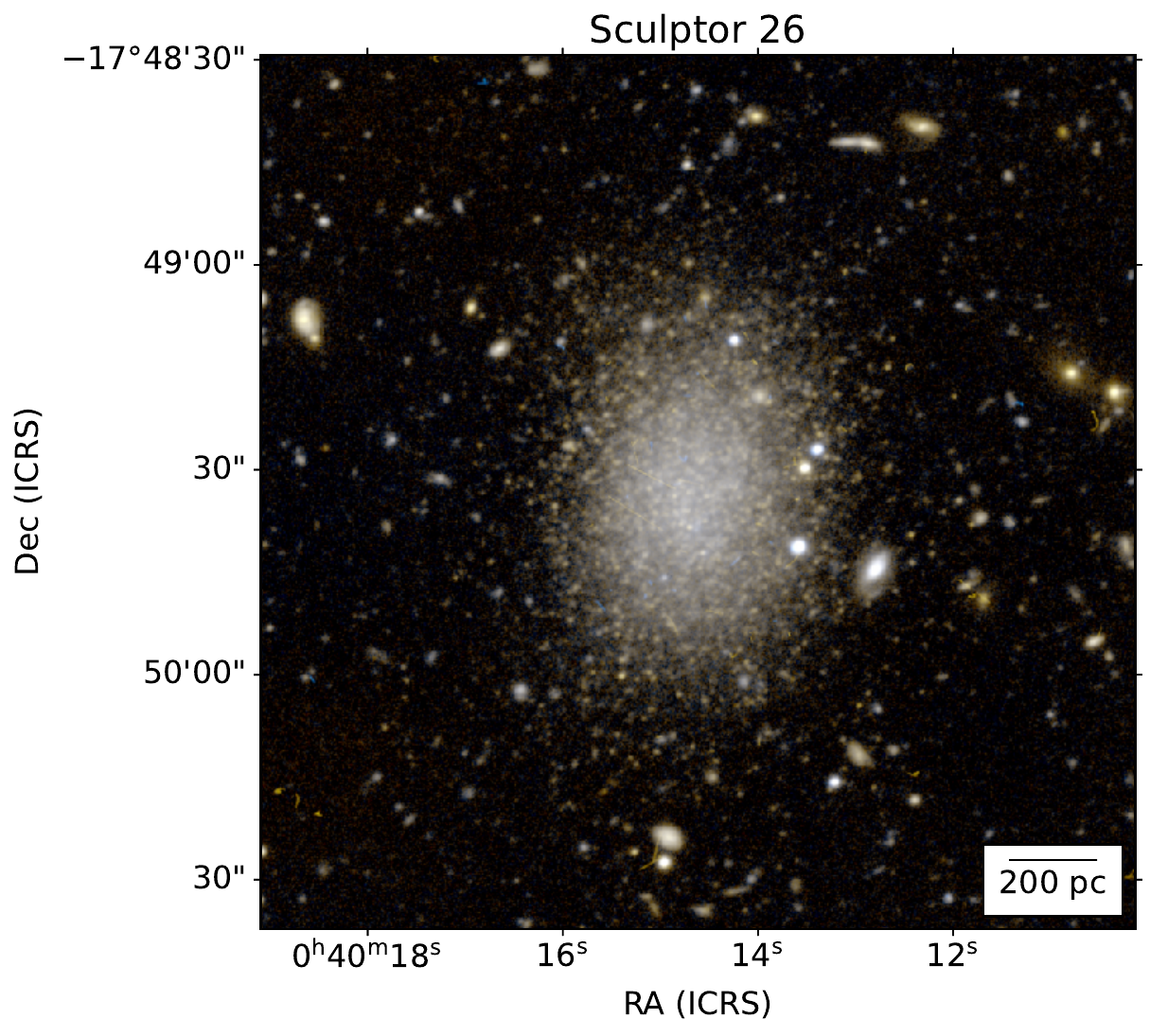}
    \includegraphics[width=1.0\columnwidth]{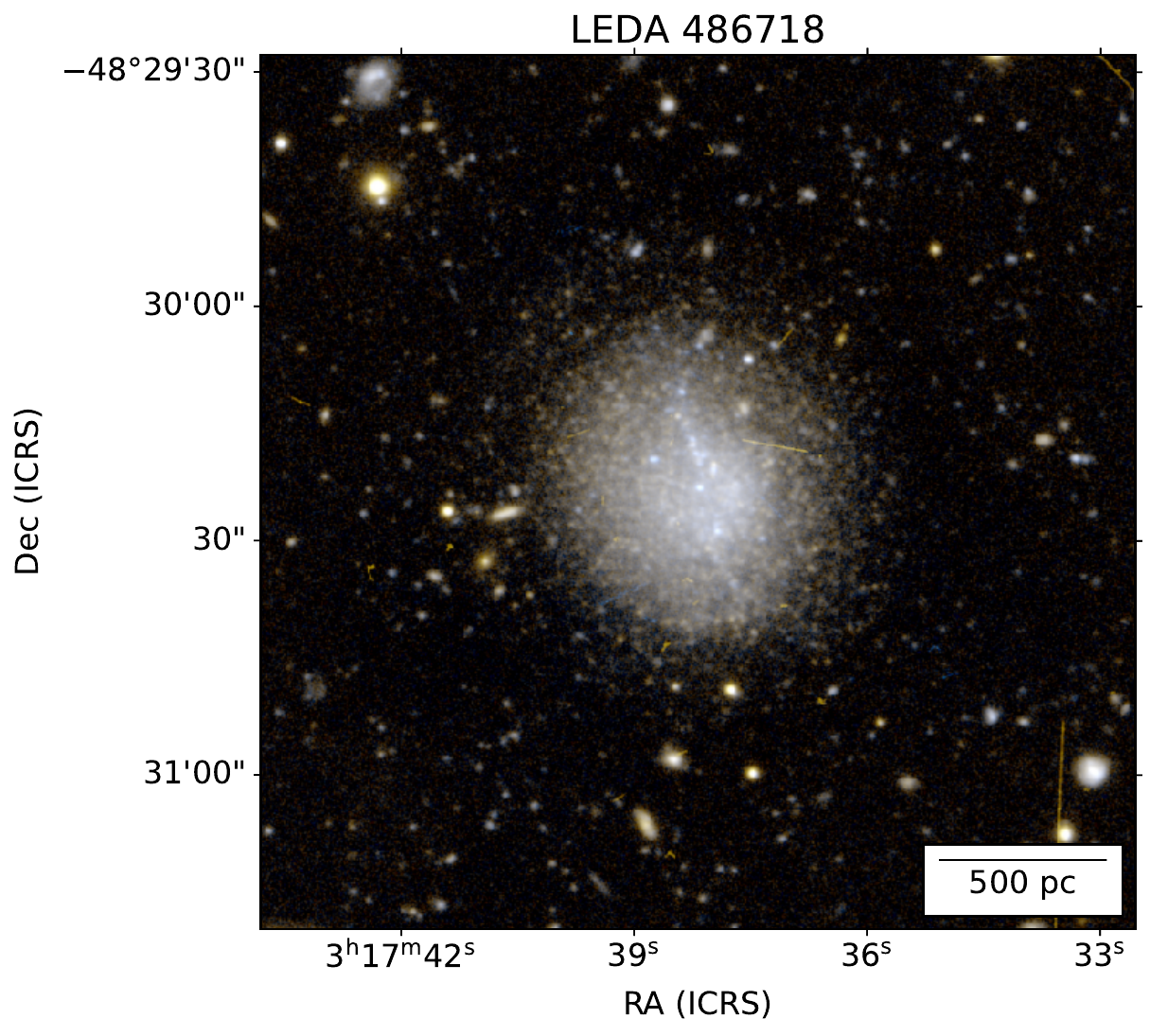}
    \caption{\textit{Top row}: Magellan/Megacam $g + r$ color images of Hydrus~A (upper left) and Cetus~B (upper right). \textit{Bottom row}: Gemini/GMOS $g + r$ color images of Sculptor~26 and LEDA~486718. All images are oriented such that north is up and east is left. These galaxies are dominated by red stellar populations, with LEDA~486718 containing some central, blue star formation. It is also evident that Cetus~B and Sculptor~26 have somewhat elongated morphologies.}
    \label{fig:optical_img}
\end{figure*}

\subsection{Swift UV Observations}
\label{subsec:swift}

All targets in this work lie within the footprint of the GALEX All-Sky Imaging Survey, with the exception of Sculptor~26. To complement this and measure star formations rates in the recent past ($\approx100-200$~Myr) for Sculptor~26 we obtained UV imaging with the Ultraviolet/Optical Telescope (UVOT; \citealt{roming2005}) on the NASA Neil Gehrels Swift Observatory \citep{gehrels2004} in the UVM2 band, which is the most similar to the GALEX NUV band \citep{hoversten2009}. Sculptor~26 was observed on October 10, 2025 with an exposure time of 1281~s. It is clearly detected in UVM2 (see \autoref{sec:appendix}). Our star formation rate calculation is described in \autoref{subsec:sfr} with the results presented in \autoref{tab:props}.

\section{Galaxy Properties}
\label{sec:analysis}

\begin{table*}[]
    \centering
    \caption{Properties of Hydrus~A, Cetus~B, Sculptor~26, and LEDA~486718}
    \begin{tabular}{r|c|c|c|c}
    \hline\hline
        Parameter & Hydrus~A & Cetus~B & Sculptor~26 & LEDA~486718 \\ \hline
        R.A. (J2000)  & 03:44:46.40 & 00:40:34.90 & 00:40:14.62 & 03:17:38.18 \\
        Dec. (J2000)  & $-$70:21:15.80 & $-$20:33:25.60 & $-$17:49:33.28 & $-$48:30:23.69  \\
        $m_{r,TRGB}$ (mag) & $24.64\pm0.07$ & $24.59\pm0.12$ & $24.52\pm0.07$ & $25.40\pm0.05$\\
         $m-M$ (mag) & $27.65\pm0.20$ & $27.60\pm0.16$ & $27.53\pm0.09$ &$28.41\pm0.08$\\
        Distance (Mpc) & $3.38^{+0.32}_{-0.30}$ & $3.32^{+0.25}_{-0.23}$ & $3.21\pm0.13$ & $4.80\pm0.17$\\
        $m_{g}$ (mag) & $18.66\pm0.02$ & $19.68\pm0.04$ & $16.61\pm0.03$ & $17.05\pm0.01$ \\
        $m_{r}$ (mag) & $17.98\pm0.05$ & $19.11\pm0.04$ & $16.05\pm0.02$ & $16.61\pm0.02$ \\
        $(g-r)$ (mag) & $0.68\pm0.05$ & $0.57\pm0.06$ & $0.56\pm0.04$ & $0.44\pm0.02$ \\
        $M_{V}$ (mag) & $-9.39\pm0.20$ & $-8.26\pm0.17$ & $-11.25\pm0.10$ & $-11.62\pm0.08$ \\
        $\log{L_{V}}$ ($L_{\odot}$) & $5.69\pm0.08$ & $5.24\pm0.07$ & $6.43\pm0.04$ & $6.58\pm0.03$\\
        $r_{h}$ (arcsec) & $11.52\pm1.2$ & $12.72\pm0.37$ & $17.99\pm0.27$ & $12.14\pm0.14$\\
        $r_{h}$ (pc) & $188^{+26}_{-25}$ & $205^{+17}_{-15}$ & $279\pm12$ & $282\pm11$ \\
        $\epsilon$ & $0.15\pm0.03$ & $0.47\pm0.02$ & $0.32\pm0.02$ & $0.20\pm0.02$\\
        $\theta$ (deg) & $0.0$ & $10.83\pm0.59$ & $-3.95\pm2.79$ & $18.74\pm0.67$ \\
        $\log{(\rm{SFR}_{NUV}/\msun \rm{yr}^{-1})}$ & $<-4.3$ & $<-4.6$ & -$3.37\pm0.13$ & $-3.74 \pm 0.03$ \\
        $\log{(\rm{SFR}_{FUV}/\msun \rm{yr}^{-1})}$ & $<-4.3$ & $<-4.8$ & & $-4.02 \pm 0.10$ \\
        $\log{M_{*}}/$\msun & $6.0\pm0.1$ & $5.4\pm0.1$ & $6.6\pm0.1$ & $6.5\pm0.1$ \\
        $\log{M_{\rm{H}\,\rm{\textsc{i}}}}/$\msun & $<6.27$ & $<6.17$ & $<6.11$ & $<6.58$\\ \hline
    \end{tabular}\\ [4pt]
    \noindent Conversions from SDSS filter band to Johnson-Cousins $V$-band is facilitated by the \citet{jordi2006} photometric conversions. $r_{h}$ is the semi-major axis of the ellipse containing half the total integrated light, not a circularized radius. The physical quantities ($r_{h}$, $\epsilon$, $\theta$) are derived from both the $g$ and $r$-bands, where we quote the average between the two. Errors are calculated as the standard deviation between the two quantities. Magnitudes are calculated within $2\times r_{h}$. Stellar masses are derived from mass-to-light ratios. See text for details on calculations. 
    \label{tab:props}
\end{table*}

\begin{figure*}
    \centering
    \includegraphics[width=\columnwidth]{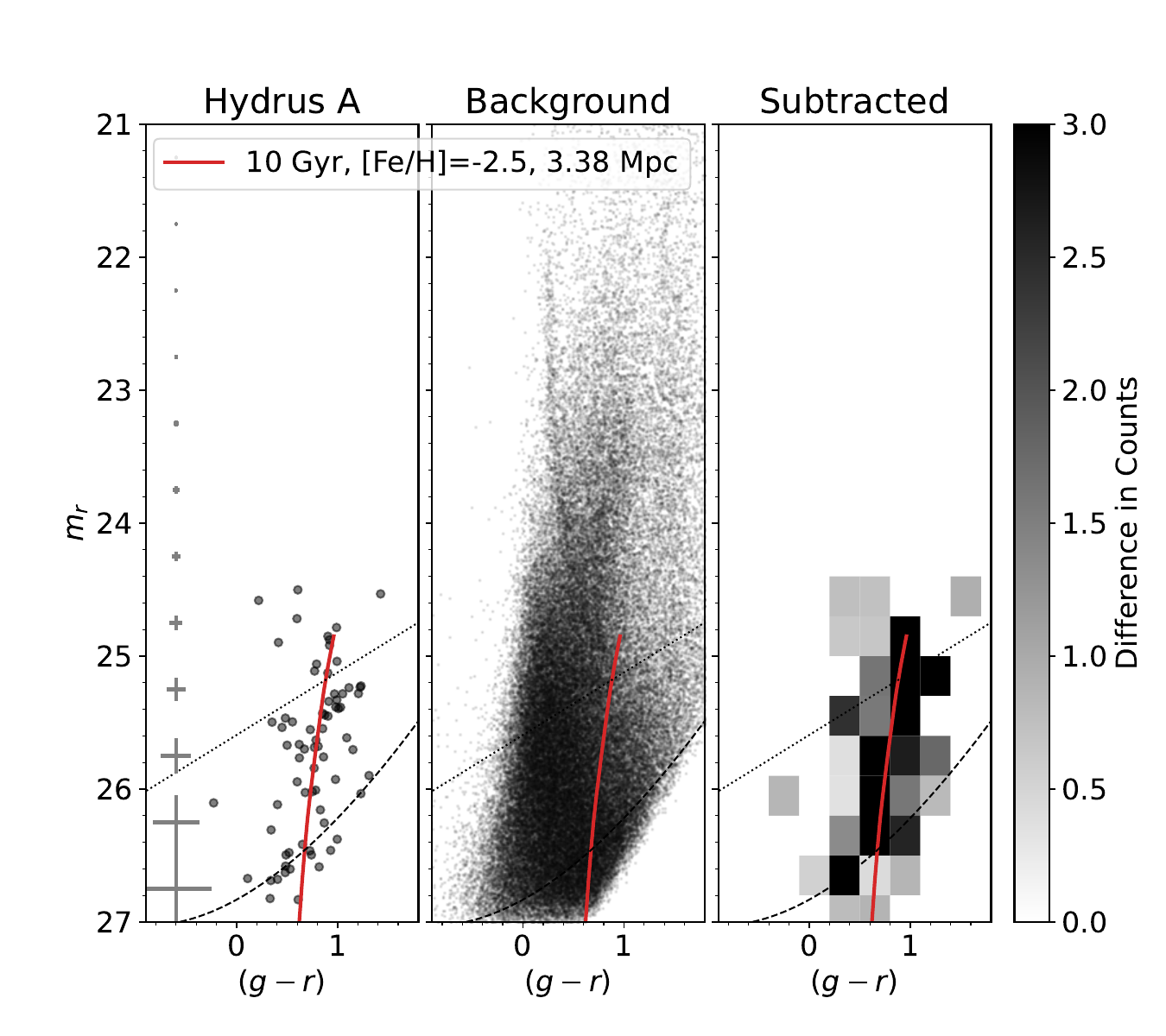} 
    \includegraphics[width=\columnwidth]{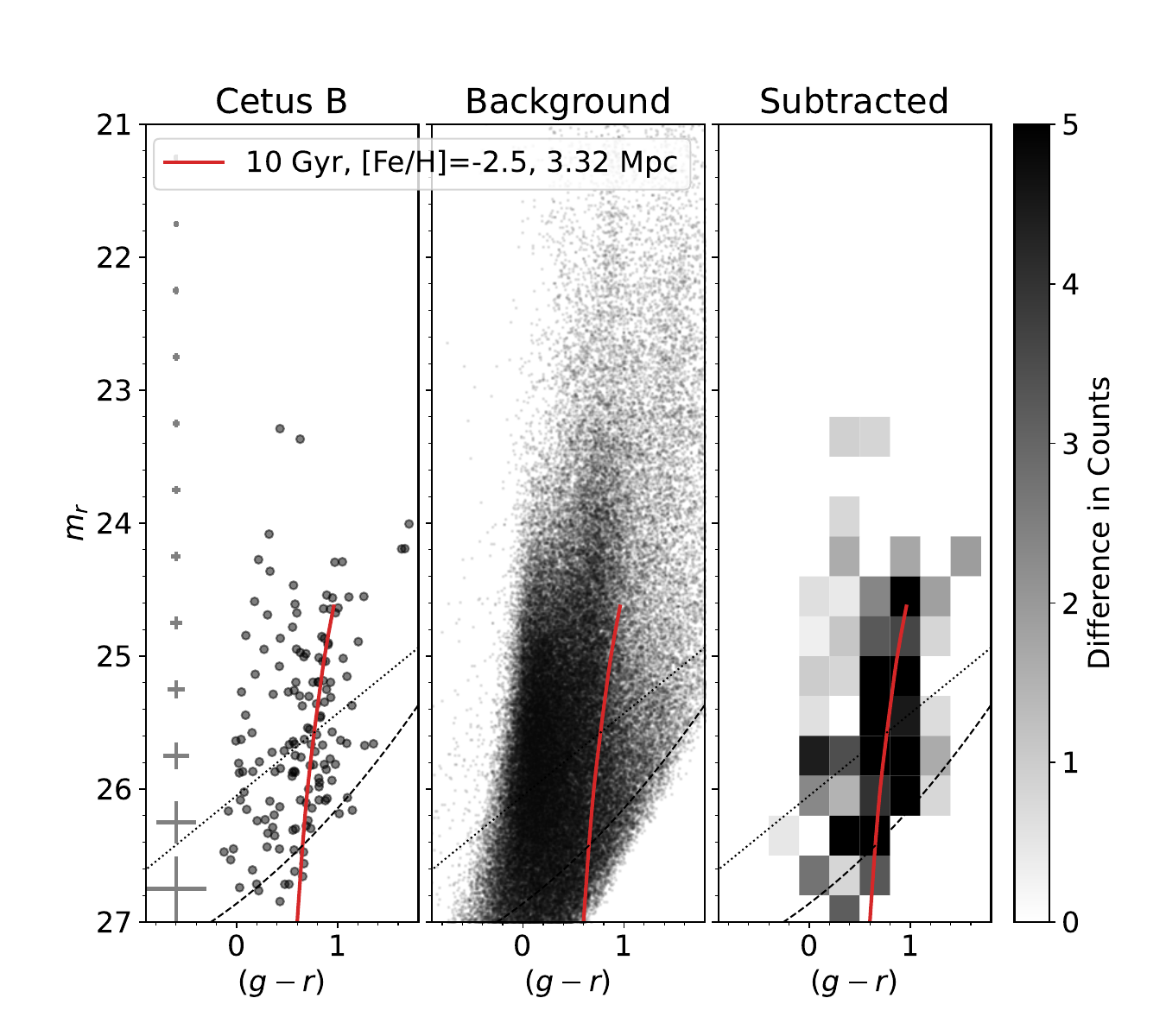} \\
    \includegraphics[width=\columnwidth]{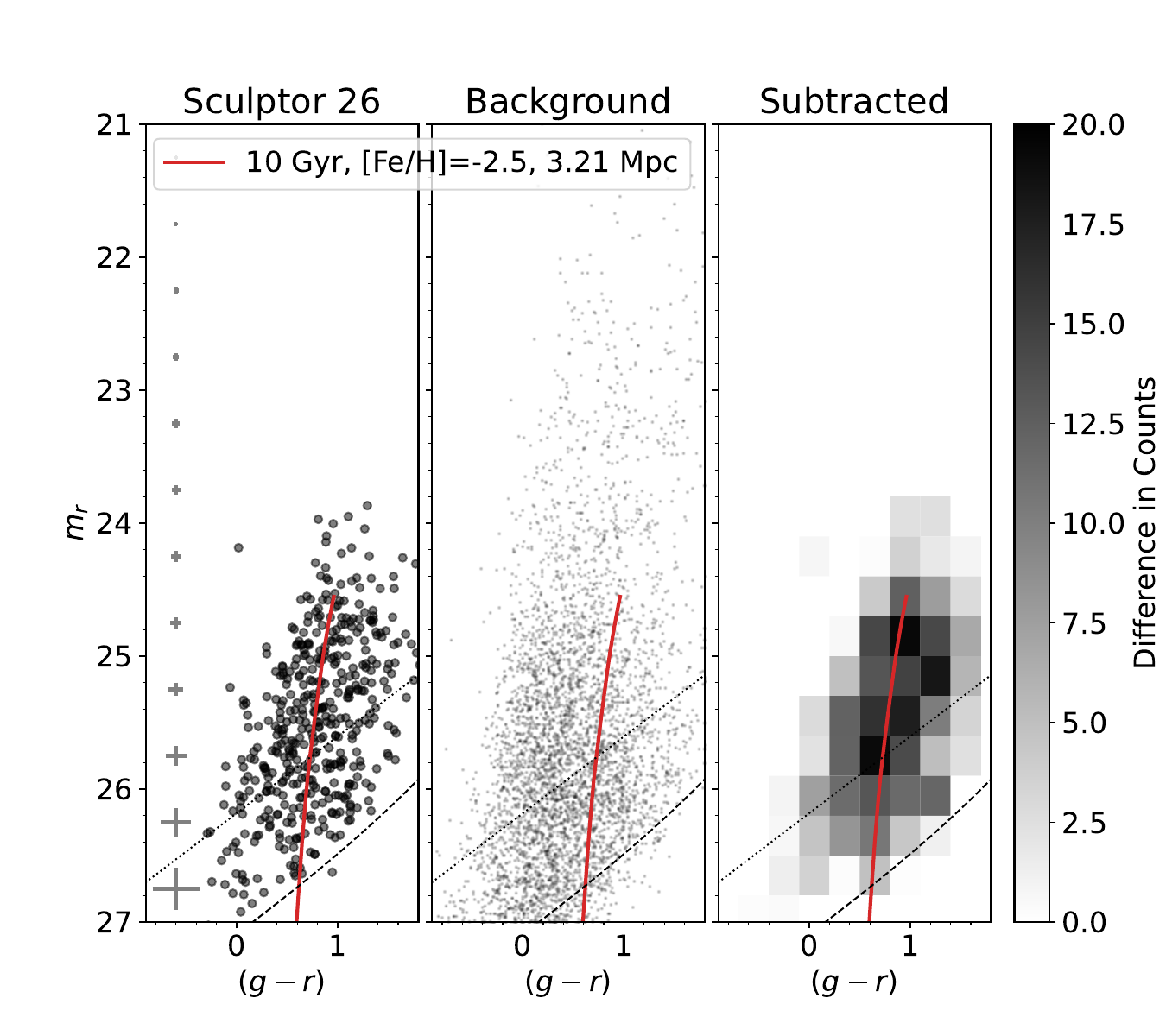} 
    \includegraphics[width=\columnwidth]{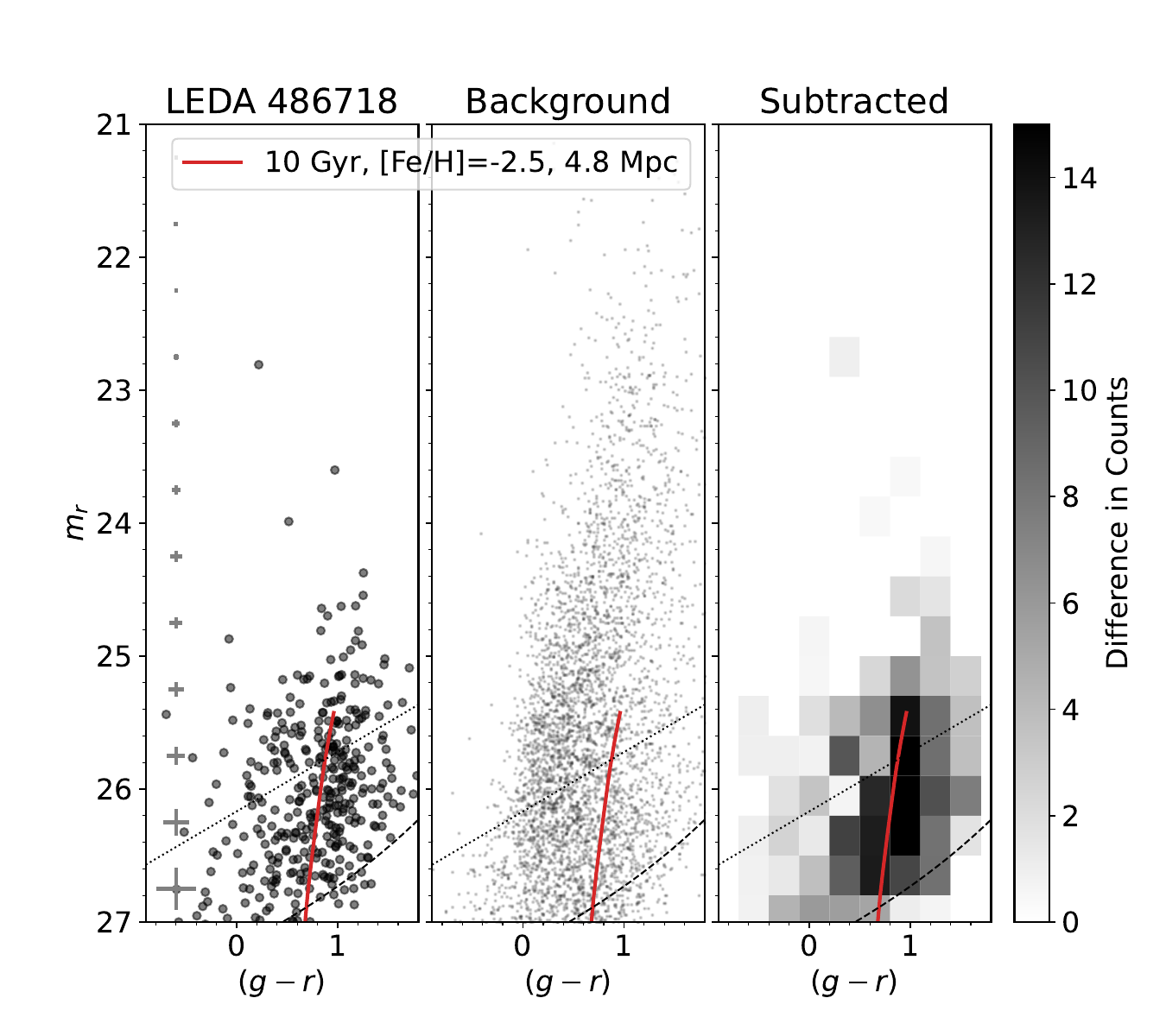} 
    \caption{Each set of 3 panels depicts the CMD of each galaxy. In each panel the dotted line marks the 90\% completeness limit and the dashed line indicates the 50\% completeness limit. The error bars on the left indicate the typical photometric uncertainties in bins of 0.5 mag in $r$-band. \textit{Left:} The CMD of stars selected within a circular aperture of the galaxy, where obvious contaminants are excluded (foreground Milky Way stars and background galaxies). \textit{Middle:} CMD of background stars in the full field of view. \textit{Right:} Binned, background-subtracted CMD. This is performed by first binning both the galaxy CMD and full background CMD. Then the binned background CMD is scaled to cover the same area as the CMD of the galaxy before the scaled background CMD is then subtracted from the galaxy CMD. In all instances the galaxy CMD clearly separates from the background CMD. Overlaid in each panel is a 10~Gyr, [Fe/H]=$-$2.5 Dartmouth isochrone \citep{dotter2008} in red, set at the TRGB distance (\autoref{tab:props}) of the respective galaxy.}
    \label{fig:CMDs}
\end{figure*}

\subsection{Color and Morphology}

Our sample, Hydrus~A, Cetus~B, Sculptor~26, and LEDA~486718, appears to be fairly red in the public DESI Legacy imaging. This is confirmed by our own deep imaging (see \autoref{fig:optical_img}) and photometric analysis in \autoref{subsec:structural_params} (see the $(g-r)$ row in \autoref{tab:props}). Integrated colors and morphologies are commonly used as a surrogate for dwarf galaxy stellar populations, with ``early-type" dwarfs typically being red, smooth, and symmetric, indicative of quenched star formation \citep[e.g.,][]{carlsten2021,greene2023}. This description broadly applies to our systems, although our deeper imaging reveals that LEDA~486718 has some resolved stars blueward of the red giant branch in the galaxy's outskirts, and hosts a small concentration of bright, blue stars near its center - though the central region is partially unresolved due to crowding. Overall, the sample remains more regular than the recently star-forming dwarfs in the SEAMLESS survey (e.g., Pavo, Corvus~A, and Kamino; \citealt{jones2023,jones2024,mutlupakdil2025}) and more similar in appearance to quenched field dwarfs such as Tucana B \citep{sand2022} and other typical dwarf spheroidals.

In addition to the global colors and morphologies, our deep imaging resolves individual stars in each galaxy (\autoref{fig:CMDs}), allowing us to examine their stellar populations directly. In all four systems we clearly detect red giant branch (RGB) stars, though in some cases only in the lower-density outskirts where crowding is minimal. Only Cetus~B does not suffer from notable central crowding. 
For LEDA~486718, crowding prevents us from resolving the central blue stars seen in the imaging, but we resolve RGB stars at larger radii, as well as a small set of bluer stars. Sculptor~26 may suffer similarly, but the presence of central young blue stars is less clear in the imaging. Hydrus~A suffers from both central crowding and contains a prominent foreground Milky Way main sequence that intersects its RGB, although this only slightly complicates the interpretation of its CMD (see next).  

\subsection{Distance}
\label{subsec:distance}

Distances are derived using the tip of the red giant branch (TRGB) method \citep{dacosta1990,lee1993,makarov2006}. With this approach the sharp break at the bright end of the old RGB population of stars is used as a standard candle. The inner regions of all of the dwarf galaxies except for Cetus~B suffer from varying degrees of crowding due to high stellar densities. As a result we manually construct circular annuli where the inner radius is set outside of where crowding is significant and the outer radius is set to two times our calculated half light radius (see \autoref{subsec:structural_params} for calculations). We then visually inspect the stars included in our annular selection and exclude obvious contaminants (e.g., very bright stars in the foreground and shredded galaxies masquerading as point sources in the background). These cuts allow us to produce clear color-magnitude diagrams (CMDs), presented in \autoref{fig:CMDs}, in order to measure the TRGB. 

We follow the procedure of \citet{crnojevic2019} for deriving TRGB distances, which we briefly summarize. The photometric uncertainty, bias, and completeness function are determined from our artificial star tests. They are then modeled by continuous functions and convolved with a model luminosity function. The resulting luminosity function is then fit via a nonlinear least-squares algorithm to the RGB of each galaxy's CMD. The TRGB values are based on the $r$-band calibration of $M_{r}^{\rm{TRGB}}=-3.01\pm0.01$ derived by \citet{sand2014}. Our derived TRGB values, distance moduli, and distances are reported in \autoref{tab:props}. The distances to all of our targets are between 3 and 5~Mpc (including uncertainties). Over-plotted in each panel of \autoref{fig:CMDs} is an old (10 Gyr), metal-poor Dartmouth isochrone \citep{dotter2008}, shifted to the derived TRGB distances. We have not attempted to assess metallicities, so the isochrones are an arbitrary old age.

Given the sparse population of the RGB for Hydrus~A and the Milky Way foreground main sequence (seen in the ``background'' panel for Hydrus~A) that directly intersects our RGB,  this distance measurement is the most tenuous. To evaluate whether the Milky Way main sequence could bias the derived distance to smaller values (so a slightly brighter TRGB magnitude), we conducted a simple test. We first estimated the expected number of foreground stars within the RGB selection region by scaling the number of stars in the background ($>3\times r_{h}$ beyond the galaxy) by the relative on-sky area, yielding $\approx4$ expected contaminants. We then simulated the effect of these contaminants by randomly removing four stars from the Hydrus~A RGB region, repeating this process 1000 times. For each of the 1000 realizations, we performed background subtraction and refit the old, metal-poor Dartmouth isochrones, varying the distance modulus from 27 to 28 mag in 0.01 mag increments. Across these simulations, the recovered distance modulus shifts by up to 0.16 mag fainter relative to our fiducial measurement, consistent with the hypothesis that Milky Way foreground stars can bias the derived distance to smaller values. We interpret the 0.16 mag shift as the scale of the uncertainty introduced by foreground contamination, not as a fixed correction. Because this represents an additional uncorrelated uncertainty, we combine it in quadrature with the statistical TRGB uncertainty (0.12 mag), resulting in a final adopted distance modulus of $27.65 \pm 0.20$ mag.

\subsection{Photometry and Structural Parameters}
\label{subsec:structural_params}

Because each galaxy is semi-resolved, and there is crowding in the central regions, we take an approach based off of that of \citet{jones2024} to measure the structural parameters of our targets. The structural parameters of each of the four dwarf galaxies are measured in the same way, which we describe below.

With the coadded images, we visually inspect to identify the approximate center, axis ratio, and position angle of each target by manually constructing an ellipse that aligns with the visible boundaries of the galaxy. Then sources that fall outside of the manually defined elliptical region around the galaxy are detected using the $\tt{photutils}$ segmentation algorithm with a $3\sigma$ threshold. This created a mask to exclude foreground stars, background galaxies, and any other contaminants that might influence our measurements. Additional contaminating sources within the galaxy aperture are manually masked.  
To account for the spatial variance of the sky, we estimate the background using $\tt{Background2D}$ in $\tt{photutils}$ with a median filter and the source mask. The background is then subtracted (note that variation across the background is quite small after data reduction). 

To measure structural parameters we then smooth the images of Hydrus~A and Cetus~B with a Gaussian kernel ($\sigma=2$) in order to aid in the detection of the low surface brightness edges of these galaxies. Smoothing is not necessary for Sculptor~26 and LEDA~486718.  The segmentation algorithm is rerun to isolate the galaxy and measure ellipticity ($\epsilon$; \autoref{tab:props}) and position angle ($\theta$; \autoref{tab:props}). We then construct concentric elliptical annuli centered on the galaxy, using the derived physical parameters. The integrated flux is measured within each successive aperture to create a cumulative light profile. Integrated flux is measured at the point that the profile reaches the sky level, and the half-light radius ($r_{h}$; \autoref{tab:props}) is measured as the semi-major axis of the ellipse containing half of the integrated light. Integrated magnitudes are derived from the total flux within the cumulative light profile, converted to magnitudes through our derived photometric zero points, and corrected for Galactic extinction using the \citet{schlafly2011} reddening maps.

Uncertainties on the flux in each band were derived empirically, accounting for both background noise and the uncertainty in the half-light radius. To quantify uncertainty associated with background noise, we randomly place a set of ``empty'' apertures (300) across the image. Each aperture was identical in size and shape to the aperture used for the galaxy flux measurement (an elliptical aperture with semi-major axis $2\times r_{h}$), but positioned randomly in regions free of bright sources, beyond $3\times r_{h}$ from the target, and away from the image edges. The fluxes measured within these apertures provide an estimate of the background noise contribution, where the standard deviation represents the corresponding flux uncertainty for an aperture area equivalent to those used to calculate our galaxy photometry.

The uncertainty on the half-light radius was estimated by using a procedure parallel to the background-noise test described above. For each elliptical annulus that contributes to the cumulative light profile, we first quantified the uncertainty in the enclosed flux due to background fluctuations. To do this, we randomly placed background apertures whose areas matched the area of that annulus, and used the standard deviation of their fluxes as the flux uncertainty at that radial step. We then propagated these flux uncertainties into an uncertainty on the half-light radius using a Monte Carlo approach: for each realization, the cumulative light profile was perturbed by drawing flux values for each annulus from Gaussian distributions centered on the measured fluxes with widths set by their corresponding uncertainties. The half-light radius was then recomputed for every realization. The standard deviation of the resulting distribution of half-light radii represents the uncertainty in $r_{h}$, capturing the impact of measurement noise and background variability on the derived size.

The total flux uncertainty is converted to magnitudes using $\sigma_{m,\rm{flux}} = \frac{2.5}{\ln{10}}\frac{\sigma_{F,\rm{total}}}{F}$. Lastly we incorporate the calibration uncertainty from the photometric zero point by adding it in quadrature. 


The same masking, background subtraction, shape measurement, surface brightness profile fitting, and integrated light analysis were repeated for both $r$- and $g$-band images. In \autoref{tab:props}, the values presented for ellipticity, position angle, and half light radius, and their corresponding errors are presented as the mean and standard error between the individual photometric band measurements. 

We compare our results to the DECaLS imaging measurements for Cetus~B from \citet{martinezdelgado2024}. The position angle that \citet{martinezdelgado2024} measured, derived instead using $\texttt{GALFIT}$ \citep{peng2002,peng2010}, is consistent with our measurement to  within the error. We find a slightly rounder ellipticity ($0.47\pm0.02$) than their $0.64\pm0.01$, but both results are fairly elliptical. Our derived half light radii are smaller and hence our derived magnitudes are slightly fainter (0.3-0.4 mag), but overall the results are not significantly different. Further calculations performed by \citet{martinezdelgado2024} assumed a distance of 3.7~Mpc for Cetus~B whereas we derive a closer distance of $3.32^{+0.25}_{-0.23}$.

\begin{figure*}
    \centering
    \includegraphics[width=0.8\linewidth]{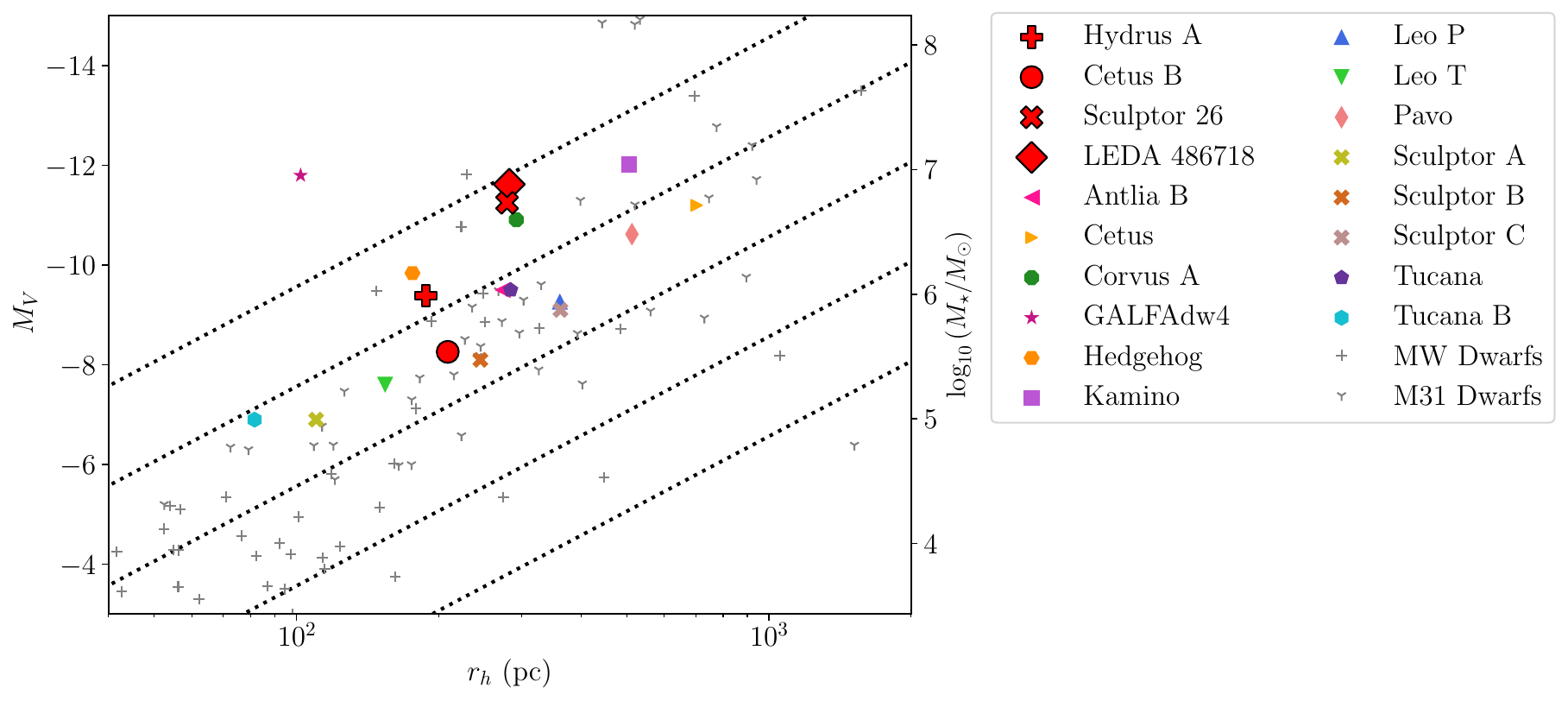}
    \caption{Size-luminosity relation of Local Volume dwarfs, with Hydrus~A, Cetus~B, Sculptor~26, and LEDA~486718 depicted with large red symbols. Diagonal dotted lines show constant surface brightness assuming an exponential profile (24, 26, 28, 30, 32 mag arcsec$^{-2}$ from bottom to top). Grey points are sourced from the Local Volume Database \citep{pace2024}. Other noteworthy Local Volume dwarfs are also highlighted \citep{irwin2007,mcconnachie2012,giovanelli2013,mcquinn2015b,sand2015b,sand2022,sand2024,bennet2022,jones2023,jones2024,li2024,mutlupakdil2025}.}
    \label{fig:size_lum}
\end{figure*}

In \autoref{fig:size_lum} we plot our sample on the size-luminosity relation compared to dwarfs in the Local Group utilizing the Local Volume Database\footnote{\url{https://github.com/apace7/local_volume_database}} \citep{pace2024} and a selection of other notable dwarfs in the Local Volume. 
Over-plotted are lines of constant surface brightness assuming an exponential profile (Sérsic index $n=1$) for surface brightness values 24, 26, 28, 30, 32 from bottom to top. All of the dwarfs in our sample lie within the scatter of the relation, but are on the higher surface brightness end, especially Sculptor~26 and LEDA~486718. However, as mentioned in \citet{jones2024}, this is likely in part the result of observational bias since high surface brightness dwarfs are easier to identify, especially outside of the Local Group. Sculptor~26 and LEDA~486718, which contain tracers of recent star formation (see \autoref{subsec:sfr}), also have correspondingly higher surface brightnesses than Hydrus~A and Cetus~B, which have no evidence of recent star formation. 

\subsection{Star Formation Rates}
\label{subsec:sfr}

Star formation rates (SFRs) for Hydrus~A, Cetus~B, and LEDA~486718 are derived from the near ultraviolet (NUV) and far ultraviolet (FUV) data within the GALEX all-sky survey \citep{galex2005}. While LEDA~486718 has a clear detection in the UV, Hydrus~A and Cetus~B do not, so we derive upper limits on their star formation rates. Sculptor~26 lies just outside the GALEX footprint, so we acquired Swift UVOT data to derive a SFR. Derived quantities are reported in \autoref{tab:props}.

We perform aperture photometry on the GALEX NUV and FUV tiles using the same process as \citet{jones2022}. We used elliptical apertures defined by twice the galaxy’s half-light radius along the major axis (our $r_{h}$	is measured along the major axis, not circularized) and the measured axis ratio, matching the apertures adopted for the optical photometry. Bright foreground sources and image artifacts are masked, and background noise uncertainties are estimated empirically by randomly placing 10,000 empty apertures across each GALEX tile, matched in area to the science aperture and restricted to regions free of extremely bright sources. Fluxes are converted to AB magnitudes using the calibrations of \citet{morrissey2007}, and corrected for Milky Way extinction adopting $R_{\rm NUV} = 8.20$ and $R_{\rm FUV} = 8.24$ from \citet{wyder2007}. De-reddened flux densities are then converted to luminosities using our measured distances to each galaxy. SFRs are calculated following the procedure of \citet{iglesias2006}, adopting a bolometric absolute magnitude of the Sun of 4.74.

Swift UVM2 photometry is derived much in the same way as our optical and GALEX photometry. Note that GALEX NUV and Swift UVM2 have very similar central wavelengths \citep[e.g.][]{cook2025}. 
Count rates within the aperture are converted to AB magnitudes using the Swift UVM2 calibration zero point\footnote{https://swift.gsfc.nasa.gov/analysis/uvot\_digest/zeropts.html} of 18.54, and then to flux densities including compensating for UVOT's sensitivity loss.
The observed magnitude and flux density is corrected for Milky Way extinction with $R_{UVM2}=6.99$ \citep{yi2023}. 
Our measured magnitude for Sculptor~26 is $m_{UVM2}=18.1$. 
SFRs are then calculated following the procedure of \citet{iglesias2006} as before.

The CMDs for our sample (\autoref{fig:CMDs}) are dominated by red, evolved stars, reflecting the fact that they probe the less crowded outer regions where the stellar populations are uniformly old. Despite this, both Sculptor~26 and LEDA~486718 have detectable star formation, suggesting the presence of younger stars concentrated in the central regions where crowding limits our CMD sensitivity. The NUV is most sensitive to star formation that occurred within the past $\approx100$~Myr, and FUV can be sensitive to even younger stars \citep[e.g.,][]{lee2009,lee2011}. For Hydrus~A, Cetus~B, and LEDA~486718, the star formation rates/limits are comparable in the NUV and FUV, meaning that the mean star formation likely has not varied substantially in the past $\sim100$~Myr. Other isolated low-mass dwarfs with extant star formation rates such as Leo P ($\log{(\rm{SFR}/\msun \rm{yr}^{-1})} = -4.4$, $\log(\rm{sSFR})=-10.1$; \citealt{mcquinn2015b}), Corvus~A ($\log{(\rm{SFR}/\msun \rm{yr}^{-1})} = -3.25\pm0.07$, $\log(\rm{sSFR})=-9.5$; \citealt{jones2024}), and Pavo ($\log{(\rm{SFR}/\msun \rm{yr}^{-1})} = -4.0^{+0.3}_{-0.8}$, $\log(\rm{sSFR})=-10.8$; \citealt{jones2023}) have comparable measured SFRs and specific star formation rates (sSFR = SFR/$M_{*}$) to Sculptor~26 ($\log(\rm{sSFR})=-10.0$) and LEDA~486718 ($\log(\rm{sSFR})=-10.2$). We note that Leo~P, Pavo, and Corvus~A are a few times lower in mass than Sculptor~26 and LEDA~486718.


\subsection{Stellar Mass and $\hi$ Mass}

We estimate the stellar masses of the four dwarf galaxies using their $(g-r)$ colors, integrated magnitudes, and the mass-to-light ratio relations of \citet{zibetti2009} and \citet{into2013}, adopting the mean of the two methods for each system. The resulting stellar masses span a range of $5.4 \lesssim \log(M_{*}/\msun) \lesssim 6.6$ (\autoref{tab:props}). Sculptor~26 is the most massive galaxy in the sample ($\log(M_{*}/\msun)=6.6\pm0.1$), closely followed by LEDA~486718 ($\log(M_{*}/\msun)=6.5\pm0.1$); both are comparable to the isolated dwarf Kamino \citep[$\log(M_{*}/\msun)=6.50^{+0.15}{-0.11}$;][]{mutlupakdil2025}. Hydrus~A is less massive ($\log(M{*}/\msun)=6.0\pm0.1$), consistent with isolated dwarfs such as Hedgehog \citep[$\log(M_{*}/\msun)=5.8\pm0.2$;][]{li2024} and Pavo \citep[$\log(M_{*}/\msun)=6.08^{+0.14}_{-0.04}$;][]{mutlupakdil2025}. Cetus~B is by far the lowest-mass system in our sample ($\log(M_{*}/\msun)=5.4\pm0.1$), similar to the extremely low-mass star-forming dwarf Leo~P \citep[$\log(M_{*}/\msun)=5.43^{+0.06}_{-0.07}$;][]{mcquinn2024a}, although Leo~P's mass is derived from a very deep CMD rather than a global color scaling relation.

We also use the derived NUV star formation rates to obtain independent, order-of-magnitude stellar mass estimates for Sculptor~26 and LEDA~486718, assuming a constant SFR over a Hubble time ($\approx1.3\times10^{10}$ yr). This yields $\log(M_{*}/\msun)=6.8\pm0.1$ for Sculptor~26 and $\log(M_{*}/\msun)=6.4\pm0.1$ for LEDA~486718 - both consistent with the mass-to-light–based results within their uncertainties.

We estimate upper limits on the \hi\ content of our targets using spectra from the \hi\ Parkes All Sky Survey (HIPASS; \citealt{barnes2001}) spectral server. We do not identify any significant emission peaks that may correspond to the neutral gas content of any of our targets. This does not completely rule out the possibility that neutral gas exists in these targets, since the HIPASS data is not particularly deep, but for now we estimate upper limits for the \hi\ masses of these systems. We assume that the dwarf radial velocities are sufficiently high enough to not be blended with the \hi\ emission of the Milky Way. To address Milky Way contamination we apply cuts to the \hi\ spectra. The spectrum for Hydrus~A is the most contaminated by the Milky Way where we apply a cut $-100 < cz_{\odot} < 400$. For Cetus~B we cut $-200 < cz_{\odot} < 100$. For Sculptor~26 and LEDA~486718 we cut $-100 < cz_{\odot} < 100$. These latter cuts are still well away from the circular velocity of NGC~253 ($\approx261~\kms$; \citealt{westmeier2017}), which Cetus~B and Sculptor~26 are near. The spectra are Hanning smoothed, with an rms noise of 8.1 mJy at a velocity resolution of 13 \kms (27 \kms\ post smoothing). We then calculate the $3\sigma$ integrated flux limit assuming a line width of 25 \kms\ for a dwarf-galaxy \hi\ line (Leo P's \hi\ velocity width is 24 \kms; \citealt{giovanelli2013}). Our values are reported in the last row of \autoref{tab:props}.

Comparing our \hi \ mass limits to other known low-mass galaxies, we would generally expect to be able to detect gas reservoirs on par with that of Corvus~A ($\log{M_{\rm{H}\,\rm{\textsc{i}}}/ \msun} = 6.55\pm0.06$; \citealt{jones2024}, \citealt{mutlupakdil2025}), but we would be unable to detect those of either Leo~P ($\log{M_{\rm{H}\,\rm{\textsc{i}}}/ \msun} = 5.91$; \citealt{giovanelli2013}, \citealt{mcquinn2024a}) or Pavo ($\log{M_{\rm{H}\,\rm{\textsc{i}}}/ \msun} = 5.79$; \citealt{jones2025}). Also, because of the distance to our four targets, the \hi \ limits are several times worse than for the quenched low-mass galaxy, Tucana B ($\log{M_{\rm{H}\,\rm{\textsc{i}}}/ \msun} < 5.6$; \citealt{sand2022}). This in turn means that only Sculptor~26 can be said to be gas-poor (i.e., $M_{\rm{H}\,\rm{\textsc{i}}}/M_\ast < 1$) with confidence.

\subsection{Environment}

\begin{figure*}
    \centering
    \includegraphics[width=1.0\columnwidth]{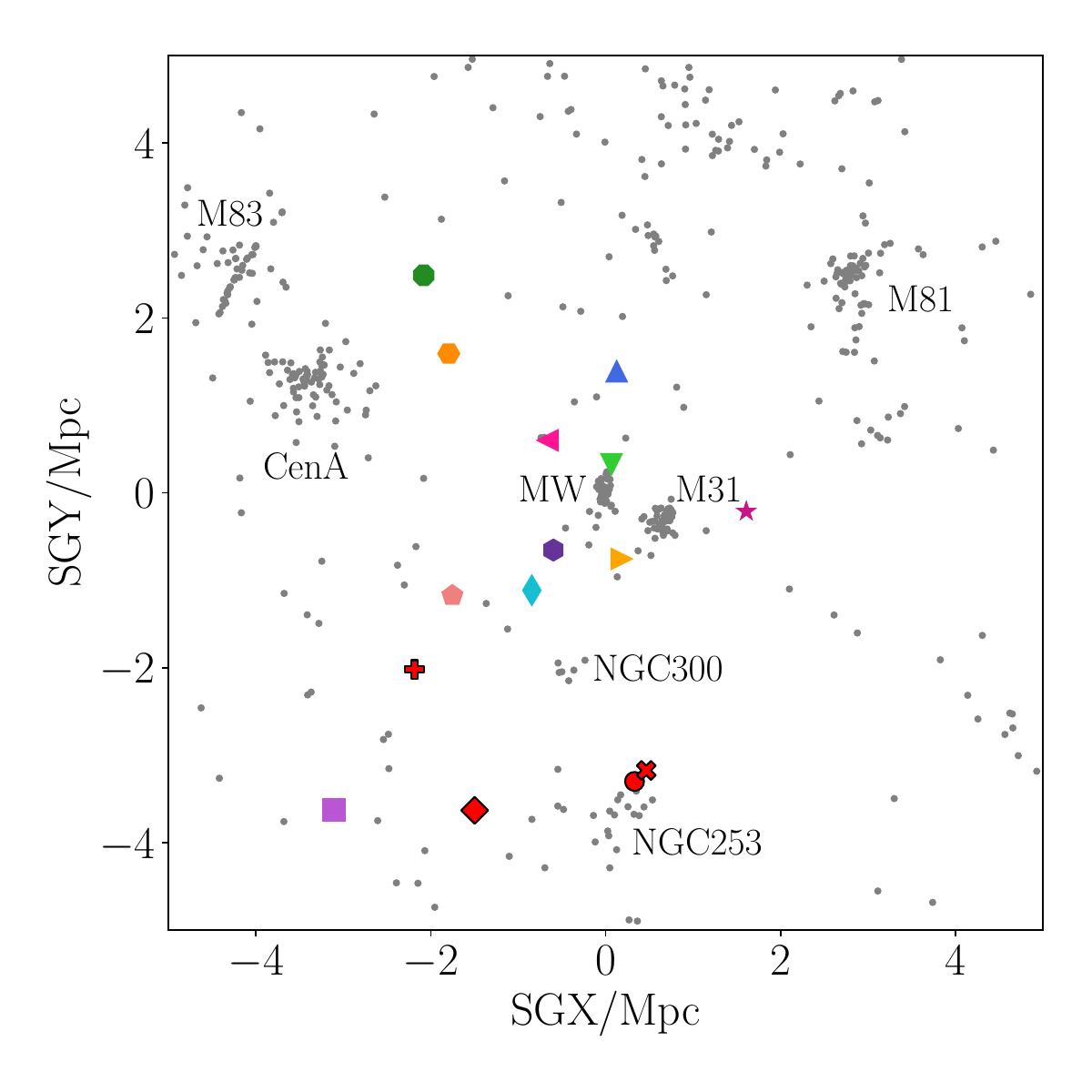}
    \includegraphics[width=1.0\columnwidth]{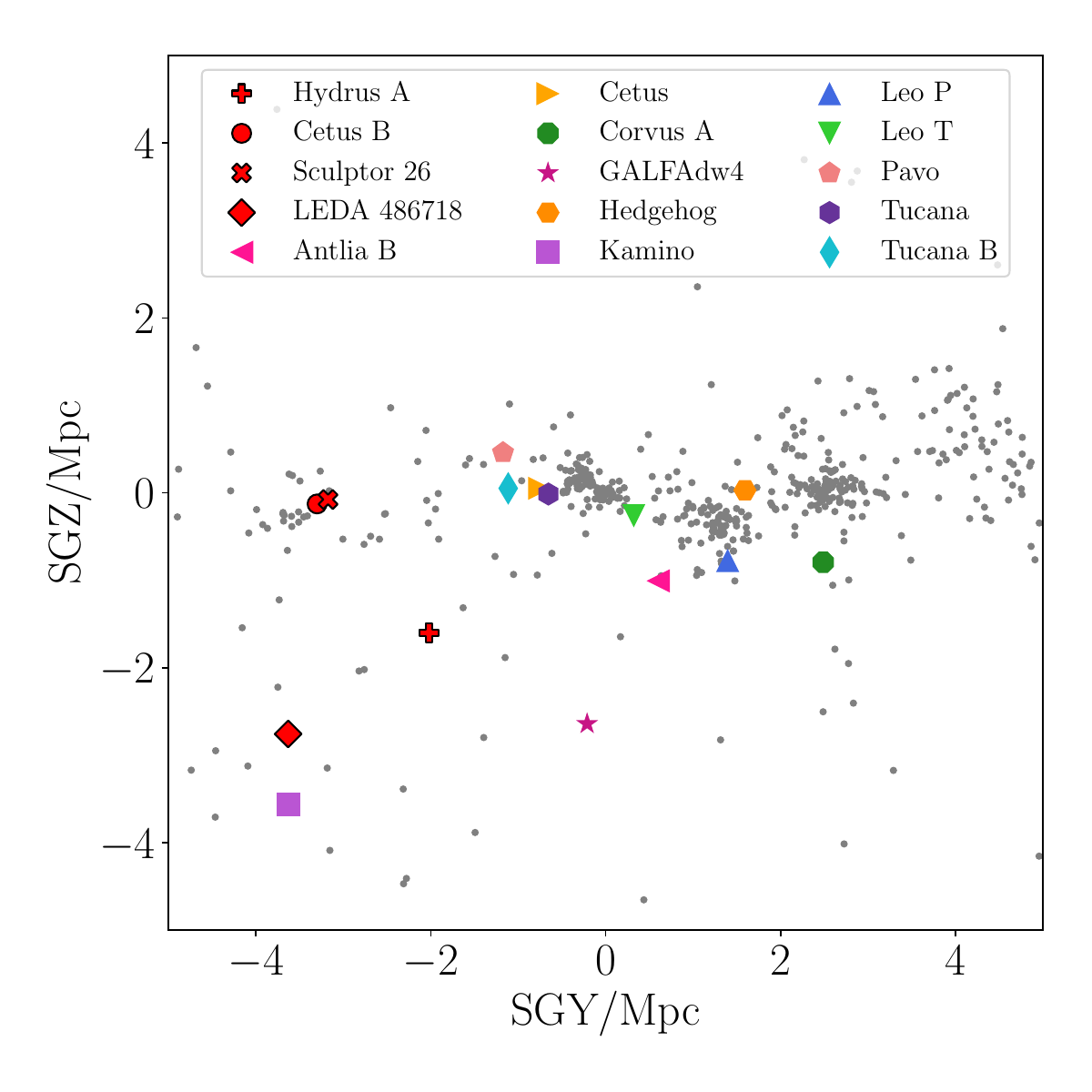}\\
    \caption{Supergalactic XY (left) and YZ (right) projections of objects in the nearby galaxy catalog \citep{karachentsev2019}. In the XY plane we label notable nearby galaxy groups. Hydrus~A, Cetus~B, Sculptor~26, and LEDA~486718 are marked with red points. Other notable nearby dwarf galaxies are included for comparison. Hydrus~A and LEDA~486718 are isolated, while Cetus~B and Sculpture~26 are in the periphery of the NGC~253 group.}
    \label{fig:environment}
\end{figure*}

In \autoref{fig:environment} we plot the super-galactic Cartesian XY and YZ projection of galaxies within the Local Volume ($<5$~Mpc) centered on the Milky Way. The galaxy catalog of \citet{karachentsev2019} is plotted in gray, with notable groups labeled. 

Our sample can be evenly split between those that are isolated and those that are plausibly associated with a group. Hydrus~A and LEDA~486718 do not fall within the local sheet and are isolated. Using the \citet{karachentsev2019} catalog, the nearest galaxy to Hydrus~A is KK~27,  902~kpc away (in 3D distance), measured via TRGB with HST in \citet{karachentsev2003}. For reference, the closest group to Hydrus~A is NGC~300, 2.2~Mpc away. The nearest galaxy to LEDA~486718 is NGC~1311, 821~kpc away, measured via TRGB with HST in \citet{jacobs2009}. The closest group to LEDA~486718 is NGC~253, 2.9~Mpc away. In comparison, even the most isolated faint dwarfs fall in close proximity to the local sheet, except for Kamino (nearest neighbor 947~kpc away; \citealt{mutlupakdil2025}) and GALFAdw4 (nearest neighbor 2.2~Mpc away; \citealt{bennet2022}). Therefore, in addition to Kamino and GALFAdw4, Hydrus~A and LEDA~486718 are some of the most isolated dwarf galaxies in the Local Volume that also fall outside of the local sheet. 

\begin{figure}
    \centering
    \includegraphics[width=\linewidth]{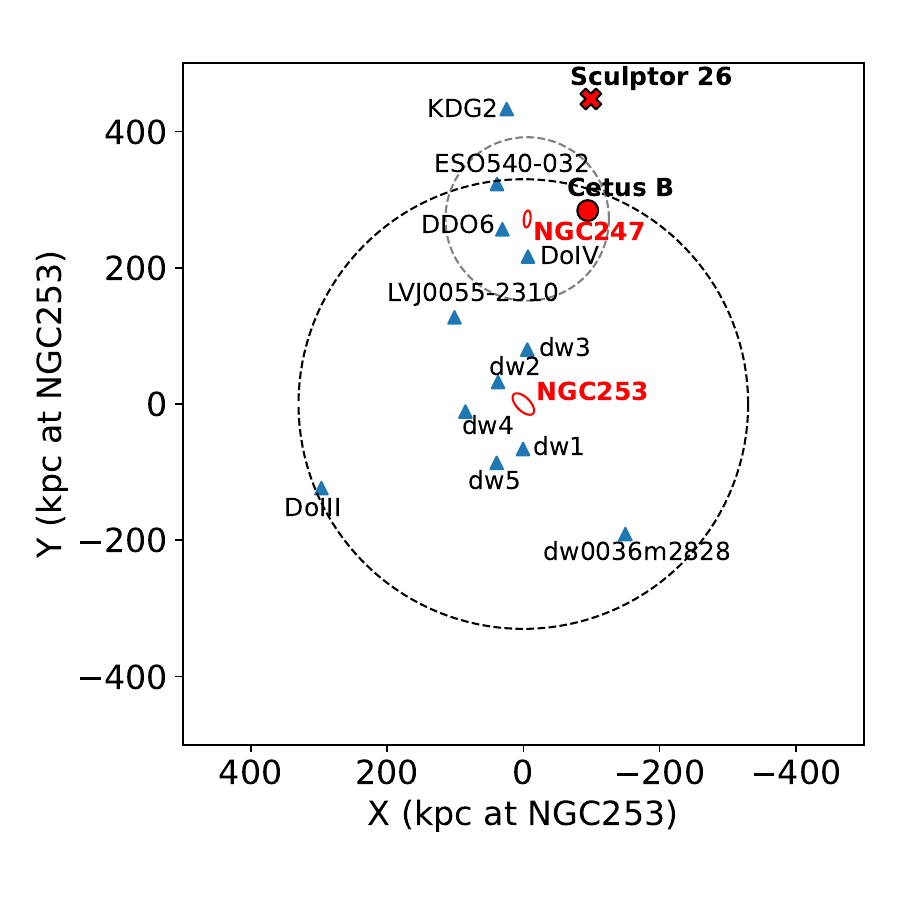}
    \caption{Known dwarf galaxies in proximity to NGC~253. The dashed black and gray circles mark the virial radii of NGC~253 and NGC~247 respectively. The sample includes dwarfs from the PISCeS survey \citep{sand2014,toloba2016,mutlupakdil2022}, and dwarfs discovered by \citet{martinezdelgado2021} (DoIII and DoIV) and \citet{carlsten2022} (dw0036m2828) with associations confirmed in \citet{mutlupakdil2024}. Cetus~B and Sculptor~26 are marked with red points. In this projection Cetus~B appears to lie within the virial radius of both NGC~253 and NGC~247. Cetus~B is within the foreground of both objects, but accounting for uncertainties on the distance may fall within the virial radius of NGC~247. Sculptor~26 is also in the foreground and within neither virial boundary.}
    \label{fig:sculptor_group}
\end{figure}

In contrast, Cetus~B and Sculptor~26 lie in proximity to the Sculptor filamentary structure and consequently the NGC~253 group. In \autoref{fig:sculptor_group} we plot Cetus~B and Sculptor~26 alongside confirmed members of the NGC~253 group. Note that in this projection Cetus~B (D $=3.32_{-0.23}^{+0.25}$~Mpc) and Sculptor~26 are in the foreground (D $=3.21\pm0.13$~Mpc) compared to NGC~253 (D $=3.5\pm0.1$~Mpc; \citealt{radburn-smith2011}), NGC~247 (D $=3.72\pm0.03$~Mpc; \citealt{jacobs2009,westmeier2017}) and even DDO6 (D $=3.44_{-0.15}^{+0.13}$ Mpc; \citealt{anand2021}). The closest galaxy to Cetus~B is DDO6, with a 3D separation of only 177~kpc measured via TRGB with HST \citep{karachentsev2003b}. The closest galaxy to Sculptor~26 is Cetus~B, separated by 190~kpc. In relation to the dominant members of the Sculptor group Cetus~B is 347~kpc away from NGC~253, and Sculptor~26 is 533~kpc from NGC~253. Assuming NGC~253 has a virial radius of approximately 330~kpc \citep{mutlupakdil2021}, Cetus~B is very close to the virial boundary (and falls within the virial radius when accounting for errors on this distance), with Sculptor~26 a bit further out. NGC~247 is a plausible sub-group of NGC~253, with a virial radius of 120~kpc \citep{mutlupakdil2021,mutlupakdil2024}.
Cetus~B lies 410~kpc from NGC~247 while Sculptor~26 lies 550~kpc away.

\section{Discussion}
\label{sec:discussion}

\subsection{Hydrus~A}

Hydrus~A is an isolated dwarf galaxy $3.38_{-0.30}^{+0.32}$~Mpc away, with a stellar mass of $M_{*} = 10^{6.04\pm0.03}$~\msun. It is the smallest galaxy in this sample ($r_{h} = 188^{+26}_{-25}$~pc), but remains consistent with the size–luminosity relation for Local Volume dwarfs (\autoref{fig:size_lum}). In physical parameter space Hydrus~A is strikingly similar to Hedgehog ($M_{*}=10^{5.8\pm0.2}$~\msun, $M_{V} = -9.84 \pm 0.16$, $r_{h} = 176 \pm 14$; \citealt{li2024}), and nearly as isolated. 
NGC~300 is the nearest group to Hydrus~A, $\sim18R_{\rm{vir}}$ away \citep{mutlupakdil2021}. 

Hydrus~A shows no sign of recent star formation. There is no UV emission to signal traces of star formation within the past $\approx100$~Myr, no detected \hi\ (although this constraint is weak), and its red optical color ($(g-r)=0.68\pm0.05$) is consistent with an exclusively old stellar population. While Hydrus~A suffers from central crowding ($<9''$), there is no evidence that there is a significant population of younger stars at the center given the red color and lack of UV detection. Its morphology is smooth and featureless, without evidence for tidal disturbances, and contains no dust lanes or similar features. Together, these properties indicate that Hydrus~A is quenched. 
Given Hydrus~A's extreme isolation, external environmental processes associated with massive hosts (e.g., ram-pressure stripping or tidal harassment) are not likely explanations for its quenched appearance. Instead, field-quenching mechanisms (e.g., filaments), cosmic quenching mechanisms (e.g., reionization), or internal quenching processes (e.g., feedback) may explain the observed properties of Hydrus~A.

Hydrus~A is low enough in mass that it falls close to the anticipated mass threshold at which galaxies are effected by cosmic reionization. Reionization is expected to be the likely cause of quenching for field dwarfs $\lesssim10^{6}\msun$ \citep[e.g.,][]{babul1992,bullock2001,benson2002,ricotti2005,weisz2014}. Very low mass galaxies are expected to be highly susceptible to gas evaporation \citep[e.g.,][]{shapiro2004,wheeler2015} and suppression of gas cooling and accretion due to the UV background from reionization \citep[e.g.,][]{okamoto2008,katz2020}. In this scenario, Hydrus~A formed its stars early and failed to subsequently accrete or retain cool gas. 

In the field filaments may also ram pressure strip the gas within a dwarf galaxy \citep{benitezllambay2013,simpson2018,benavides2025}. However, it is unclear how large of a role this mechanism plays in the quenching of field dwarfs. Given how isolated Hydrus~A is, far from the local sheet structure (see \autoref{fig:environment}), Hydrus~A would have had to pass through such a structure at much earlier times.  

Internal quenching mechanisms may also be a driver in how Hydrus~A looks today. Low-mass dwarf galaxies are more susceptible to the feedback effects of star formation and supernovae. While star formation feedback is thought to not be capable of completely quenching a dwarf galaxy, supernova feedback may be strong enough to expel most of the gas within a dwarf galaxy, especially if working in tandem with cosmic reionization \citep[e.g.,][]{salvadori2009,benitezllambay2015,jeon2017,rey2020}. For example, \citet{gallart2021} finds that nearby ultrafaint dwarf Eridanus II ($M_{*}=10^{5.05}\msun$) could be quenched by supernova feedback alone. Distinguishing internal feedback from reionization in Hydrus~A requires a detailed star-formation history derived from deep space-based optical and infra-red photometry. In the former case star formation would be expected to rapidly cease $>12$~Gyr ago, while in the latter case a more protracted SFH would be expected, potentially extending several Gyrs after the end of reionization.

A backsplash origin is technically possible, a scenario in which Hydrus~A formerly traversed a larger halo and was ejected to its current location. However, the absence of a candidate host make this scenario improbable and we can likely exclude an ancient flyby with NGC~300 given their large separation. Following the logic of \citet{li2024} if we assume an ejection velocity of $\sqrt{2}v_{circ}$ and a maximum circular velocity for NGC~300 of $\sim100~\kms$ \citep{westmeier2011}, it would take $\sim22$~Gyr to reach its current location, or longer than the age of the Universe. This is further supported by recent simulation work \citep[e.g.,][]{benavides2025} that shows that only $\sim6$\% of quenched field dwarfs are expected to be true backsplash systems under stringent isolation criteria. 
Therefore, for Hydrus~A we favor a quenching mechanism that involves either early reionization/feedback‐driven gas loss, or removal of gas via a filament/void encounter rather than a classical satellite‐host interaction. 

\subsection{LEDA~486718}

LEDA~486718 is an isolated dwarf galaxy $4.80\pm0.17$~Mpc away, with a stellar mass of $M_{*} = 10^{6.53\pm0.06}$~\msun. LEDA~486718 is isolated, and over $\sim8.8R_{\rm{vir}}$ from the nearest group, NGC~253.
Such separations strongly disfavor a recent interaction with another galaxy. 

LEDA~486718 is the largest ($r_{h} = 282\pm11$~pc; only slightly larger than Sculptor~26) and the brightest galaxy in our sample ($M_{V}=-11.62\pm0.08$). While it falls within the expectation of the size-luminosity relation, it is on the high surface brightness end. In this physical parameter space both LEDA~486718 and Sculptor~26 are very similar to Corvus~A ($M_{*}=10^{6.20^{+0.15}_{-0.07}}~\msun$, $M_{V} = -10.91\pm0.10$, $r_{h} = 292 \pm 16$~pc; \citealt{mutlupakdil2025}), but are redder in appearance.

UV emission indicates star formation within the past $\approx100$~Myr for LEDA~486718, consistent with its blue color ($(g-r)=0.44\pm0.02$; the bluest in our sample). The central region appears to have bright blue stars that are not fully represented in our CMD due to central crowding, but are plausibly the site of recent star formation. LEDA~486718 is redder than Corvus~A or Pavo (both $(g-r)=0.3$; \citealt{jones2023,jones2024}), but the CMD of LEDA~486718 still clearly displays both an old RGB sequence and some bluer, younger stars. While less prominent than in Corvus~A, Pavo, or Leo~P \citep[see ground-based CMD in][]{mcquinn2013}, LEDA 486718 is the only galaxy in our sample with an identifiable younger stellar population  distinct from background contaminants (see \autoref{fig:CMDs}). This young component, combined with measurable UV SFR, suggests that LEDA 486718 has experienced star formation in the past $\sim$100 Myr.

Unlike the isolated star-forming dwarf Leo~P, LEDA 486718 does not have a detected \hi\ reservoir. The non-detection could indicate that (1) the remaining gas is below current sensitivity limits of the HIPASS survey, (2) the \hi\ could overlap with the Milky Way ($-100 < cz_{\odot} < 100$), (3) the gas is predominantly ionized, or (4) the recent star-formation episode exhausted or expelled the neutral gas component. Such low or absent \hi\ content in otherwise star-forming (until recently) dwarfs has been noted in other isolated systems, for example Pavo \citep{jones2023,jones2025}. 


Standard galaxy formation theory suggests that low-mass dwarfs commonly form stars in bursty, intermittent episodes driven by internal feedback through star formation \citep[10-100~Myr timescales, e.g.,][]{hopkins2014,fitts2017,emami2019} or supernovae \citep[$>100$~Myr timescales, e.g.,][]{stinson2007,bland-hawthorn2015,wheeler2019,hirai2024}. In either scenario gas is disrupted and heated, typically flowing out into a hot halo where it cools until star formation is able to continue. Observational \citep[e.g.,][]{giovanelli2015} and theoretical works \citep{rey2020,rey2022} suggest that the very low mass, star forming galaxies may even periodically become invisible to \hi\ surveys due to these feedback processes (assuming the \hi\ is not detectable with deeper observations). LEDA 486718 may be undergoing such a cycle, where the gas reservoir is low-mass, difficult to detect at HIPASS sensitivity, and intermittently converted into stars. 



The combination of isolation and clear evidence of recent star formation suggests that LEDA~486718 is an example of a low-mass field dwarf with either low level star formation, or internally regulated, intermittent star formation. Deeper \hi\ observations and photometric observations to measure the star formation history are necessary.

\subsection{Cetus~B}

Cetus~B is a very low mass $M_{*} = 10^{5.40\pm0.04}$~\msun, quenched dwarf galaxy, located $3.32^{+0.25}_{-0.23}$~Mpc away. Cetus~B is the faintest ($M_{V} = -8.26\pm0.17$) and least massive galaxy in our sample. In terms of targets outside of the Local Group, Cetus~B is the most similar to Sculptor~B, a quenched galaxy which may plausibly be an extreme backsplash galaxy of NGC~300 \citep{sand2024}. 

Cetus~B has no tracers of star formation in the past $\approx100$~Myr in the UV and a red stellar population ($(g-r)=0.57\pm0.06$). These properties suggest that Cetus~B contains a predominantly old stellar population and its star formation has quenched, but we note that the \hi\ limits are insufficient to draw conclusions about gas in the system. Cetus~B has no dust lanes or detectable star-forming regions, and appears elongated in the north-east southwest directions. 

Cetus~B is not isolated, $\sim1.1R_{\rm{vir}}$ from NGC~253, and $\sim3.4R_{\rm{vir}}$ from NGC~247 \citep{mutlupakdil2021}
Cetus~B is in the foreground of the projection plotted in \autoref{fig:sculptor_group}, but including error on the distance we measure to Cetus~B, it may be within the virial radius of NGC~253.

Given Cetus~B's proximity to NGC~253, coupled with elongated morphology suggestive of environmental processing, Cetus~B is a satellite/backsplash candidate of the Sculptor group. 
In the backsplash picture, Cetus~B would have passed within the virial radius of NGC~253 (or possibly NGC~247), where environmental processing - e.g. ram-pressure stripping or tidal interactions - could have substantially removed its gas supply and perturbed its stellar structure \citep[e.g.][]{slater2014,simpson2018}. If Cetus~B is a satellite of the system (and may very well be, since within error is $<1R_{\rm{vir}}$ from NGC~253) then this process may be ongoing. 


Cetus~B may also have experienced a variety of other processes (e.g., cosmic filament interactions, reionization, internal feedback, see Hydrus~A discussion) and is only coincidentally near the Sculptor group, but its close proximity to NGC~253 and distorted morphology are suggestive of a backsplash/satellite candidate (although the major axis of Cetus~B does not point towards NGC~253). Deep kinematics, gas-sensitivity-limited observations, and orbital modeling will ultimately be required to distinguish these evolutionary histories.


\subsection{Sculptor~26}

Sculptor~26 is a star-forming dwarf galaxy located at $D=3.21\pm0.13$~Mpc in the direction of the Sculptor filament, with a stellar mass of $10^{6.58\pm0.04}$~\msun, the most massive in our sample. In size and luminosity Sculptor~26 is fairly high surface brightness, closely resembling the isolated star-forming dwarfs LEDA~486718 and Corvus~A \citep{mutlupakdil2025,jones2024}. 

Like LEDA~486718, Sculptor~26 shows UV emission indicative of star formation within the past $\approx100$~Myr. The CMD of Sculptor~26 lacks an obvious young main-sequence component, although we may be missing some due to central crowding. Its morphology is smooth, without clear dust lanes or knotty star-forming regions. 
Together, these properties point to a system at the boundary between actively star forming and quenched.

Sculptor~26 does not have a detected \hi\ reservoir. However, unlike the rest of the sample, the \hi\ limit is below the stellar mass estimate, so we can infer that even if \hi\ is present then Sculptor~26 is gas poor ($M_{\rm{H}\,\rm{\textsc{i}}}/M_\ast < 1$). As we mention for LEDA~486718 the gas could just be below the sensitivity limit of HIPASS, overlap with the Milky Way in its spectrum, be predominantly ionized, or the \hi\ may have been recently exhausted. Group infall provides a fifth pathway: Sculptor~26 may be experiencing gas removal via tidal or ram-pressure interactions.

Sculptor 26 is not isolated, $\sim1.6R_{\rm{vir}}$ from NGC~253 and $\sim4.6R_{\rm{vir}}$ from NGC~247 \citep{mutlupakdil2021}.
Although in the foreground relative to NGC~253, Sculptor~26's position lies in the regime where association with the group is plausible 
\citep{fillingham2018,simpson2018}.

While Sculptor~26 is a backsplash candidate of the Sculptor group at this radius, its relatively regular morphology and nonzero recent star formation make a backsplash origin less compelling, as backsplash dwarfs are typically more strongly processed \citep{slater2014,fillingham2015,wetzel2015,fillingham2019}.
A simpler explanation is that Sculptor~26 is on first infall into the NGC~253 system and is beginning to experience environmental suppression (gas stripping takes a few Gyr after infall to MW-mass hosts). Alternatively, Sculptor~26 could be like LEDA~486718, with the UV flux reflecting either continuous or intermittent star formation. The proximity to NGC~253 leads us to favor early group processing as an explanation for Sculptor~26's current appearance, but fully distinguishing between these scenarios requires deeper \hi\ measurements, kinematic measurements to determine group association, and resolved space-based photometric star formation history measurements.

\section{Summary and Conclusion}
\label{sec:conclusion}

The environmental diversity and properties of these four systems provides an opportunity to assess how isolation and group association influence the properties of low-mass galaxies beyond the Local Group. Hydrus~A and LEDA~486718, which lie outside the local sheet, are currently among the most isolated dwarfs known in the Local Volume, while Cetus~B and Sculptor~26 occupy the periphery of the Sculptor filament and may be  associated with the NGC 253 group. Hydrus~A and Cetus~B appear to be quenched, while Sculptor~26 and LEDA~486718 harbor evidence of star formation in the past $\sim$100~Myr. None of these targets have \hi\ detections in the HIPASS survey; the \hi\ constraints on Sculptor~26 are strong enough for us to infer that it is gas poor. 

Our sample is low mass ($M_{*} \lesssim 10^7,M_\odot$), below the mass threshold where nearly all isolated dwarfs are expected to be star forming \citep{geha2012}. At this mass scale the quenched fraction of isolated galaxies rises to $\sim15$–20\% in simulations \citep{christensen2024}. Thus, the presence of both star-forming and quenched dwarfs in isolated environments is not unexpected. Hydrus~A’s low stellar mass, compact structure, red stellar population, lack of indicators of recent star formation, and extreme isolation (nearest neighbor $\sim$900 kpc) make it a candidate for quenching by reionization, the cosmic web, or self-regulating internal feedback. Conversely, LEDA~486718, with recent star formation, may represent a low-mass dwarf with either continued and steady star formation, or intermittent, and possibly bursty star formation.

The other half of our sample lies plausibly within the dynamical reach of the NGC~253 group, with both Cetus~B and Sculptor~26 located at $\lesssim2R_{\rm vir}$ of NGC~253. Yet these two systems are distinct. Cetus~B is a faint, low-mass, morphologically disturbed galaxy, with its proximity to the Sculptor group making it a backsplash or satellite candidate. If confirmed as a backsplash system, it would be among the lowest-mass backsplash systems known, comparable to Sculptor~B near NGC 300, and would serve as a valuable case study of environmental transformation around a Milky-Way-like system. Sculptor~26, by contrast, appears to be an intermediate-state dwarf: currently forming stars but red, plausibly gas poor. 
Its properties place it between isolated star-forming dwarfs and backsplash satellites, consistent with early‐stage environmental transformation. 


Overall, these four galaxies illustrate that low‐mass dwarfs outside the Local Group occupy a wide range of evolutionary states. Their gas content, recent star-formation histories, and morphological properties do not correlate in a simple way with environment. Instead, the sample supports a picture in which quenching mechanisms at low mass are diverse: internal regulation may dominate in some isolated systems; reionization or the cosmic web may suppress star formation in the most weakly bound halos; and environmentally driven processes — ram pressure stripping or tidal heating or stripping — operate near massive hosts.

Future observations are needed to clarify the evolutionary histories of these systems.
Deep \hi\ observations would determine whether weak gas reservoirs exist, potentially revealing ongoing environmentally driven stripping or re-accretion. Systemic velocities will be critical for determining whether Cetus~B and Sculptor~26 are bound to NGC~253, thereby constraining their orbital histories and backsplash likelihoods. Space-based imaging with HST or JWST to the oldest main sequence turn off would yield detailed star-formation histories, differentiating early quenching (e.g., by reionization) from more recent environmental or internal shutdown \citep[e.g.,][]{weisz2019,mutlupakdil2025}. Such data would also resolve whether LEDA~486718 has experienced bursty star formation (like Leo~P) or more steady low-level activity (as was revealed for Pavo). Together, these observations would place strong constraints on the dominance of internal versus environmental quenching channels in the lowest-mass galaxies. 

To date, SEAMLESS has identified dozens of strong dwarf–galaxy candidates across the Local Volume and has now delivered detailed follow-up characterization for seven systems \citep[see][]{jones2023,jones2024,mutlupakdil2025}. The four galaxies presented here highlight the survey’s ability to uncover both extremely isolated dwarfs and systems at the edges of group environments, populations that have been historically underrepresented in census studies. SEAMLESS will provide an unprecedented, machine-learning–enabled map of the low-mass galaxy population within 5~Mpc. Many of these additional candidates will require vetting with deeper imaging: roughly half appear to lie beyond 5~Mpc and instead resemble the SHIELD galaxies (Survey of \hi\ in Low-mass Dwarfs; \citealt{cannon2011}).  These ‘beyond-Local Volume’ systems illustrate both the power of SEAMLESS to identify low-mass galaxies and the need for careful distance confirmation. The resulting statistical sample — paired with targeted follow-up spectroscopy, \hi\ mapping, and space-based imaging — will enable transformative tests of low-mass galaxy formation, quenching pathways, and environmental processing outside the Local Group.


\begin{acknowledgments}

This paper includes data gathered with the 6.5 meter Magellan Telescopes located at Las Campanas Observatory, Chile. 

Based on observations obtained at the international Gemini Observatory (under Fast Turnaround program GS-2024B-FT-208), a program of NSF NOIRLab, which is managed by the Association of Universities for Research in Astronomy (AURA) under a cooperative agreement with the U.S. National Science Foundation on behalf of the Gemini Observatory partnership: the U.S. National Science Foundation (United States), National Research Council (Canada), Agencia Nacional de Investigaci\'{o}n y Desarrollo (Chile), Ministerio de Ciencia, Tecnolog\'{i}a e Innovaci\'{o}n (Argentina), Minist\'{e}rio da Ci\^{e}ncia, Tecnologia, Inova\c{c}\~{o}es e Comunica\c{c}\~{o}es (Brazil), and Korea Astronomy and Space Science Institute (Republic of Korea).

This work used images from the Dark Energy Camera Legacy Survey (DECaLS; Proposal ID 2014B-0404; PIs: David Schlegel and Arjun Dey). Full acknowledgment is at https://www.legacysurvey.org/acknowledgment/.

We acknowledge the use of public data from the Swift data archive.

BMP acknowledges support from NSF grant AST2508745. 
DJS \& DZ acknowledge support from NSF grants AST-2508746. 
DC acknowledges support from NSF grant AST-2508747.
DZ and RD acknowledge support from NSF AST-2006785 and NASA ADAP 80NSSC23K0471 for their work on the SMUDGes pipeline. KS acknowledges support from the Natural Sciences and Engineering Research Council of Canada (NSERC).

\end{acknowledgments}

%

\vspace{5mm}
\facilities{Magellan/Megacam, Gemini/GMOS, Swift/UVOT, GALEX}


\software{
    Astropy \citep{astropy2013,astropy2018,astropy2022}, 
    Numpy \citep{numpy},
    Scipy \citep{2020SciPy-NMeth,scipy_14593523},
    Pandas \citep{pandas_13819579},
    Photutils \citep{bradley2022},
    \texttt{DS9} \citep{DS9},
    \texttt{GALFIT} \citep{peng2002,peng2010},
    \texttt{daophot} and \texttt{allframe} \citep{stetson1987,stetson1994},
    {\sc scamp} \citep{scamp},
    DRAGONS \citep{labrie2023,labrie2023b}
}


\appendix
\section{UV Detections}
\label{sec:appendix}
In this section we present the UV images of Sculptor~26 and LEDA~486718. Sculptor~26 is constructed from the UVM2 data we acquired through Swift, as described in \autoref{subsec:swift}. LEDA~486718 is constructed from archival GALEX all sky imaging survey data \citep{galex2005}. 

\begin{figure*}
    \centering
    \includegraphics[width=0.49\textwidth]{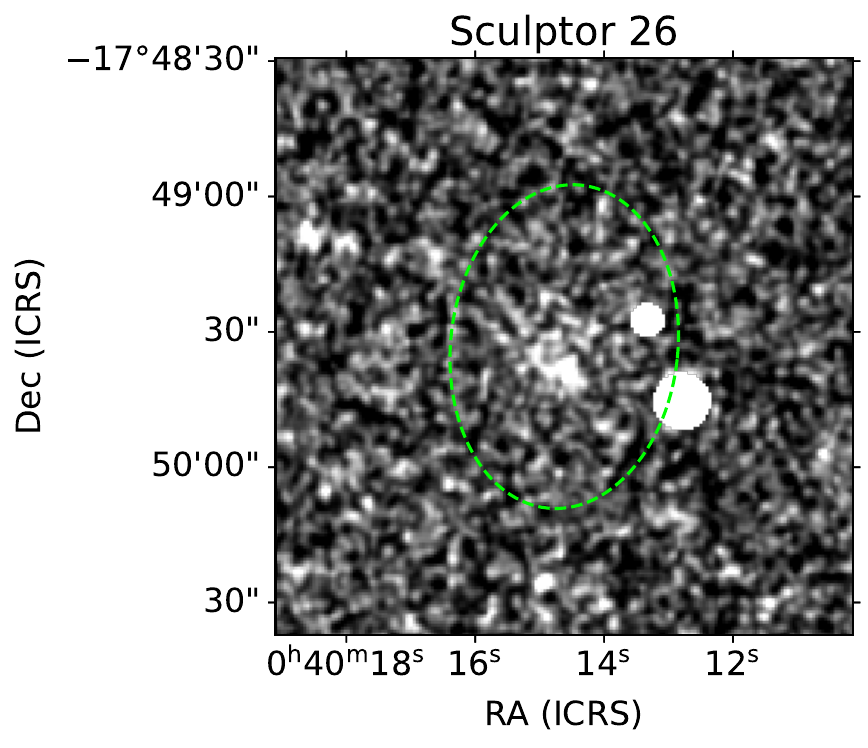}
    \includegraphics[width=0.49\textwidth]{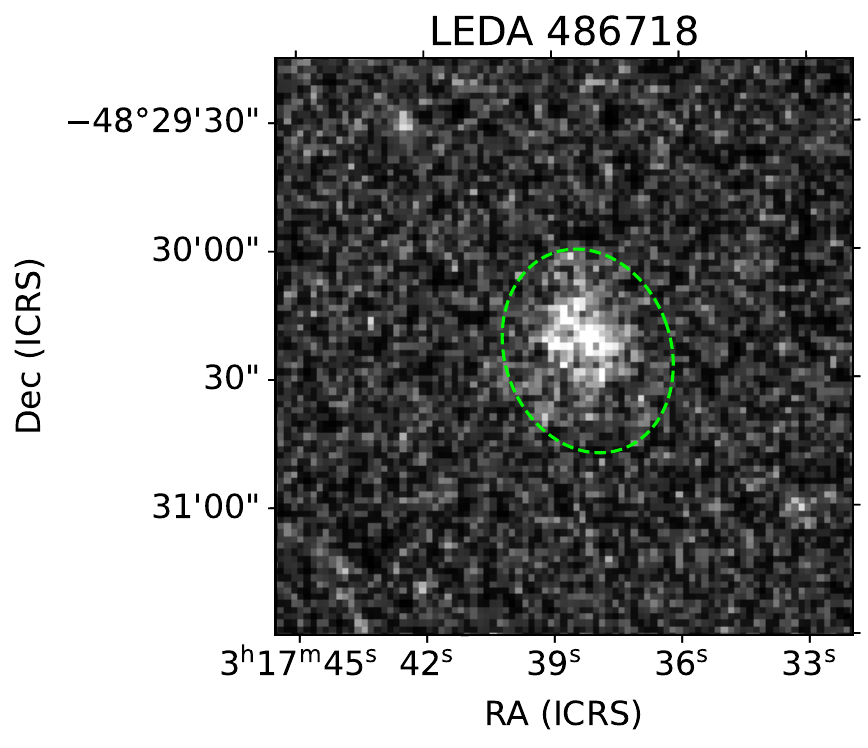}
    \caption{\textit{Left}: Swift UVM2 image of Sculptor~26. \textit{Right:} GALEX NUV image of LEDA~486718. Both images are oriented such that north is up and east is left. The dashed ellipses mark $2\times$ the optical half light radius of the respective galaxy, where the semi major axis is $2\times r_{h}$ in \autoref{tab:props} and the remaining properties of the ellipse come from the ellipticity $\epsilon$ and position angle $\theta$ of the same table. Sculptor~26 and LEDA~486718 are both clearly detected in the UV.}
    \label{fig:uv_img}
\end{figure*}


\bibliography{refs}{}
\bibliographystyle{aasjournal}



\end{document}